\documentclass[reprint,prl,aps,superscriptaddress]{revtex4-1}
\usepackage{times}
\usepackage{graphicx}
\usepackage{amsmath, braket, amsfonts}
\usepackage{amssymb}
\usepackage{natbib}
\usepackage{sidecap}
\usepackage{bm, color, ulem}
\usepackage{xcolor}
\usepackage{ragged2e}
\usepackage{wrapfig,lipsum,booktabs}
\usepackage{siunitx}

\usepackage{bibunits}
\defaultbibliography{references}
\defaultbibliographystyle{custom}

\newcommand{\abs}[1]{\left\lvert #1 \right\rvert}
\newcommand\dif{\mathop{}\!\mathrm{d}}
\newcommand{\be}{\begin{equation}}
\newcommand{\ee}{\end{equation}}

\DeclareUnicodeCharacter{03BD}{{$\nu$}}

\makeatletter
\def\maketitle{
\@author@finish
\title@column\titleblock@produce
\suppressfloats[t]}
\makeatother

\begin{document}
%TC:ignore
\title{Nanoscale electrostatic control in ultra clean van der Waals heterostructures by local anodic oxidation of graphite gates}
%\title{quasi-particle tunneling in quantum Hall point contacts \\ via anodic oxidation and transfer of graphite gates}

\author{Liam A. Cohen}
\thanks{These authors contributed equally to this work}
\affiliation{Department of Physics, University of California at Santa Barbara, Santa Barbara CA 93106, USA}
\author{Noah L. Samuelson}
\thanks{These authors contributed equally to this work}
\affiliation{Department of Physics, University of California at Santa Barbara, Santa Barbara CA 93106, USA}
\author{Taige Wang}
\affiliation{Department of Physics, University of California, Berkeley, California 94720, USA}
\affiliation{Materials Sciences Division, Lawrence Berkeley National Laboratory, Berkeley, California 94720, USA}
\author{Kai Klocke}
\affiliation{Department of Physics, University of California, Berkeley, California 94720, USA}
\author{Cian C. Reeves}
\affiliation{Department of Physics, University of California at Santa Barbara, Santa Barbara CA 93106, USA}
\author{Takashi Taniguchi}
\affiliation{International Center for Materials Nanoarchitectonics,
National Institute for Materials Science,  1-1 Namiki, Tsukuba 305-0044, Japan}
\author{Kenji Watanabe}
\affiliation{Research Center for Functional Materials,
National Institute for Materials Science, 1-1 Namiki, Tsukuba 305-0044, Japan}
\author{Sagar Vijay}
\affiliation{Department of Physics, University of California at Santa Barbara, Santa Barbara CA 93106, USA}
\author{Michael P. Zaletel}
\affiliation{Department of Physics, University of California, Berkeley, California 94720, USA}
\affiliation{Materials Sciences Division, Lawrence Berkeley National Laboratory, Berkeley, California 94720, USA}
\author{Andrea F. Young}
\email{andrea@physics.ucsb.edu}
\affiliation{Department of Physics, University of California at Santa Barbara, Santa Barbara CA 93106, USA}

\date{\today}

 \begin{abstract}
In an all-van der Waals heterostructure, the active layer, gate dielectrics and gate electrodes are assembled from two-dimensional crystals that have a low density of atomic defects. This design allows two-dimensional electron systems with very low disorder to be created, particularly in heterostructures where the active layer also has intrinsically low disorder, such as crystalline graphene layers or metal dichalcogenide heterobilayers. A key missing ingredient has been nanoscale electrostatic control, with existing methods for fabricated local gates typically introducing unwanted contamination. Here we describe a resist-free local anodic oxidation process for patterning sub 100nm features in graphite gates, and their subsequent integration into an all-van der Waals heterostructure. We define a quantum point contact in the fractional quantum Hall regime as a benchmark device and observe signatures of chiral Luttinger liquid behaviour, indicating an absence of extrinsic scattering centres in the vicinity of the point contact. In the integer quantum Hall regime, we demonstrate in situ control of the edge confinement potential, a key requirement for the precision control of chiral edge states. This technique may enable the fabrication of devices capable of single anyon control and coherent edge-state interferometry in the fractional quantum Hall regime.
\end{abstract}
\maketitle
%TC:endignore

%\begin{bibunit}[custom]

Van der Waals heterostructures have recently emerged as a rich platform to study the physics of delicate correlated electronic states, including (but not limited to) fractional quantum Hall phases\cite{zibrov_tunable_2017,li_even_2017,dean_fractional_2020}, exciton condensates\cite{dean_fractional_2020,li_excitonic_2017,liu_quantum_2017,ma_strongly_2021}, quantized anomalous Hall insulators\cite{serlin_intrinsic_2020,li_quantum_2021},  fractional Chern insulators\cite{spanton_observation_2018,xie_fractional_2021}, and superconductors\cite{cao_unconventional_2018,yankowitz_tuning_2019,zhou_superconductivity_2021,zhou_isospin_2022,zhang_spin-orbit_2022}. 
A key driver of continued improvement in sample quality has been the removal of charged impurities, first with the use of high purity two-dimensional crystals of hexagonal boron nitride (hBN) as a substrate\cite{dean_boron_2010} and gate dielectric\cite{young_electronic_2012,mayorov_direct_2011,wang_one-dimensional_2013}, and more recently with the use of graphite, rather than amorphous metal, for the gate layers\cite{zibrov_tunable_2017}. 
These `all-van der Waals heterostructures' take advantage of the fact that none of the components in the stack host dangling bonds in their two-dimensional bulk.  In addition, numerous van der Waals interfaces appear to be self-cleaning\cite{haigh_cross-sectional_2012}, irreversibly expelling hydrocarbon residues during processing and leaving an atomically uniform interface. 

A central feature of these platforms is electrostatic tunability, enabling a variety of correlation-driven ground states to be accessed by field effect gating in a single device. Electrostatic control on the nanoscale, then, allows one- and zero-dimensional structures to be created within a correlated two-dimensional state, opening the door to experiments that probe the structure of interfaces between distinct phases as well as adiabatic manipulation of individual quasi-particles and edge modes.  
A wide class of these experiments require electrostatic confinement on length scales comparable to the correlation length of superconductors or fractional quantum Hall states, which is typically below \SI{100}{nm}. 
The confining potentials are also required to be energetically uniform, in the sense that they should not introduce uncontrolled local electrical potentials larger than the $\approx \SI{1}{meV}$ energy gaps of the correlated states to be studied.  There are two options empowered by traditional electron beam lithography, which is capable of patterning at the length scales required for constructing such nano-scale potentials. 
First, the all-van der Waals geometry may be abandoned, patterning at least some gates from evaporated metal. 
Second, heterostructures may be assembled and then graphite gates patterned by subtractive processes. 
However, both techniques lead to disorder in critical regions of the device. 
For example, edge state interferometers manufactured using either technique remain limited to the integer quantum Hall regime despite the presence of well formed fractional quantum Hall phases in the two-dimensional sample bulk \cite{ronen_aharonov-bohm_2021, zimmermann_tunable_2017, deprez_tunable_2021, zhao_graphene-based_2022}.

 % The utility of developing such low-disorder, nanoscale electrostatic potentials is not restricted to the quantum Hall regime. 
 % As but one example, recent proposals for manufacturing topological superconductivity in twisted bilayer graphene nanowires require a clean gate-defined interface between superconductivity and two inter-valley coherent states \cite{thomson_gate-defined_2021}.

% Local electrostatic control is of particular interest at high magnetic fields, where clean graphene systems host a range of topologically ordered states\cite{dean_fractional_2020}.  Manipulating edge modes at electrostatically defined boundaries between topologically ordered states, for example at a quantum point contact or within an edge state interferometer, can reveal information about the quasi-particle charge and statistics \cite{nayak_non-abelian_2008}. 

% Devices fabricated in this manner exhibit rich cascades of fractional quantum Hall states including a host of states not previously observed in III-V semiconductor quantum wells in double layer systems\cite{liu_interlayer_2019,li_pairing_2019,dean_fractional_2020}, as well as even denominator states in bilayer graphene whose elementary charged excitations are predicted to obey non-Abelian statistics and may provide a promising platform for topologically protected quantum information processing \cite{nayak_non-abelian_2008, papic_topological_2014, zibrov_tunable_2017,li_even_2017,bonderson_individually_nodate}. 

\begin{figure*}
    \centering
    \includegraphics[width = 180mm]{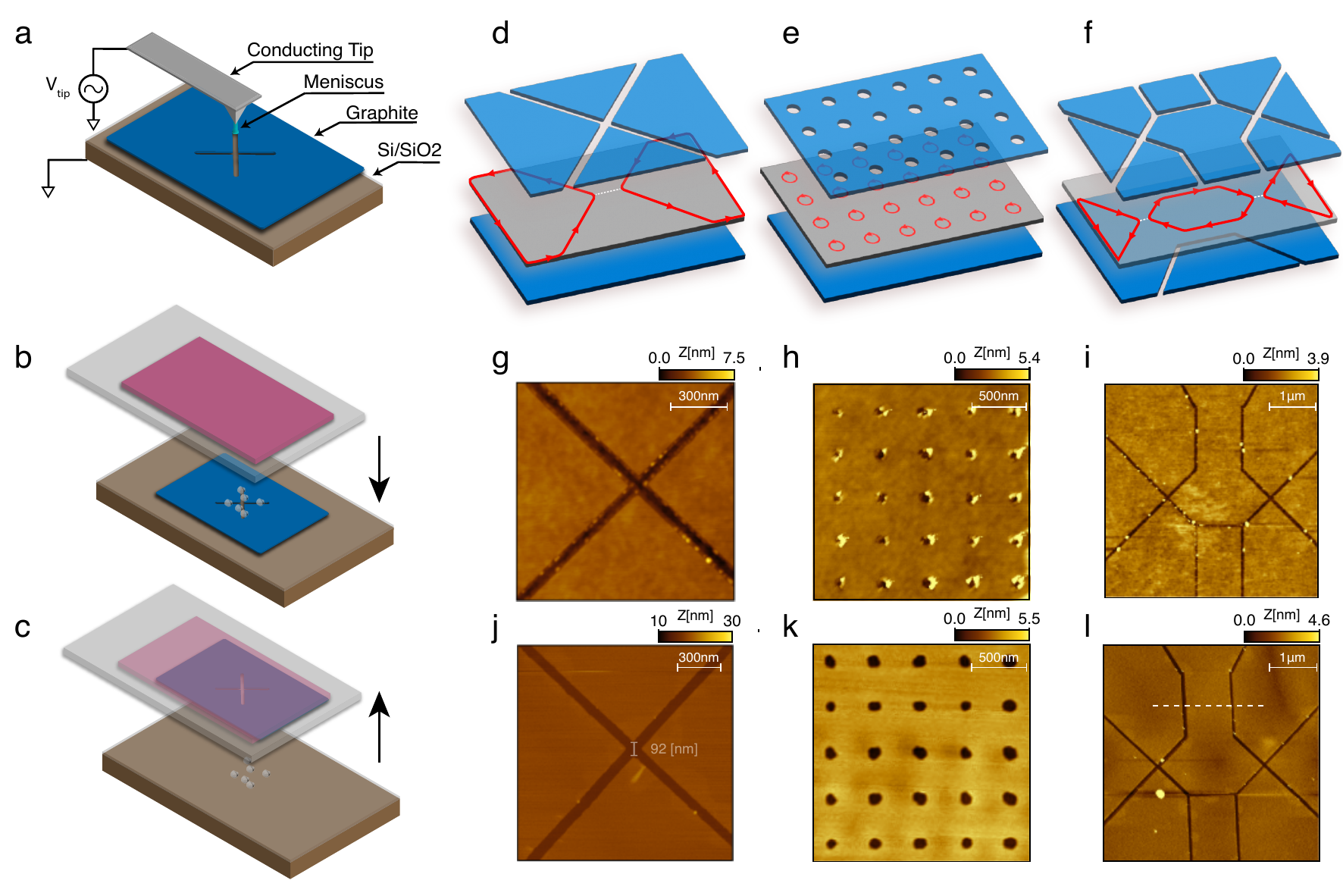}
    \caption{\textbf{Local anodic oxidation and integration of graphite gates into van der Waals heterostructures.} \textbf{(a)} A schematic example of an exfoliated graphite flake etched using atomic force microscope-actuated local anodic oxidation (AFM-LAO). Here we show the x-shaped geometry used to form the quantum point contact gates which constitute the device studied in this work.
    \textbf{(b)} A transfer stamp with an already picked-up hBN flake is engaged with the etched graphite flake which has some residual oxide residue from the AFM-LAO process.
    \textbf{(c)} Once the etched graphite is laminated onto to the hBN flake the structure can be disengaged, leaving behind the oxidized residue and resulting in a pristine microstructure. \textbf{(d)} Example edge state configuration of a quantum point contact resulting from the electrostatic potential created by an x-shaped graphite gate. \textbf{(e)} Same as panel (d) but with an array of holes in the top graphite gate used to create a superlattice of quantum dots. \textbf{(f)} Same as panels (d-e) but with an edge-state interferometer structure.  Here the isolated Fabry-P\'erot cavity can be realized by careful alignment of etched graphite gates in both the top and bottom layers of the heterostructure.  \textbf{(g-i)} AFM topography images of patterns etched into graphite flakes using AFM-LAO corresponding to the top gate structures in panels (d-f). \textbf{(j-l)} AFM topogrophy images of etched graphite on hBN post van der Waals pick-up of the flakes in panels (g-i) demonstrating that most of the oxidized residue (white dots) seen in panel (g-i) is not transferred.  
    }
    \label{fig:fab_geometry}
\end{figure*}

\begin{figure}[ht!]
    \centering
    \includegraphics[width = 88mm]{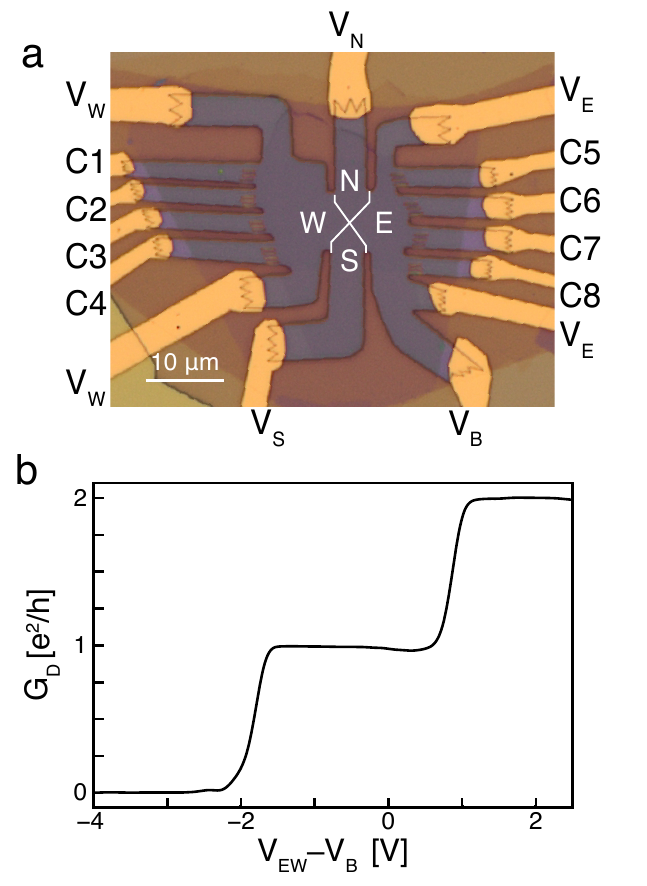}
    \caption{\textbf{Quantum point contact operation in the integer quantum Hall regime.} 
    \textbf{(a)} Optical micrograph of measured device.  Transport contacts are denoted by C1-C8.  Gate contacts are labeled by $V_i$ with an appropriate subscript to denote which region the gate controls. \textbf{b)} Diagonal conductance, $G_{\mathrm{D}}$, versus $V_{\text{EW}} - V_{\mathrm{B}}$ demonstrating integer QH edge mode partitioning at the QPC.  Here the east/west gates are swept simultaneously with the back gate such that the bulk filling factor, $\nu_{EW} \propto V_{\text{EW}} + \alpha V_{\mathrm{B}}$, remains constant.  Additionally, $\nu_{NS} \propto V_{\text{NS}} + \alpha V_{\mathrm{B}} = 0$ is held constant.}
    \label{fig:integer_qpc}
\end{figure}

\section{Device Fabrication}

Here we show how high-quality mesoscopic devices may be created by abandoning traditional lithographic techniques in favor of patterning graphite gates at sub-\SI{100}{nm} length scales using a resist-free process which takes advantage of atomic force microscope-actuated local anodic oxidation (AFM-LAO) of graphite \cite{li_electrode-free_2018}. These gates are then integrated directly into a van der Waals heterostructure using a low-strain variation of the standard dry-transfer process \cite{wang_one-dimensional_2013,
novoselov_2d_2016, masubuchi_fabrication_2009}.  Fig.~\ref{fig:fab_geometry}a-c portray a schematic description of this processs. Fabrication begins with the use of an atomic force microscope (AFM) to locally oxidize\cite{masubuchi_fabrication_2009,li_electrode-free_2018} a region of the graphite flake.  In this process, a conductive AFM tip is brought close to the graphite surface in a humid environment.  Capillary forces form a nano-scale water meniscus\cite{butt_using_2006} connecting the tip and graphite surface.  When a high frequency excitation is applied to the tip, the voltage drop across the water meniscus catalyzes oxidation of the graphite into gaseous and amorphous byproducts.  Scanning the tip across the graphite surface while this reaction occurs allows nano-scale subtractive lithography\cite{li_electrode-free_2018} without introducing contaminants to the two dimensional graphite bulk, as would occur in solvent-based resist removal processes.  The patterned graphite gate can then be integrated into a van der Waals heterostructure through pick-up by an unpatterned van der Waals flake (see Methods and Fig.~\ref{fig:fab_geometry}b-c) to produce a wide array of geometries, several examples of which are depicted in Figs. \ref{fig:fab_geometry}d-f.  

In the anodic oxidation process, amorphous residue--likely carbon and carbon oxides--typically remains, manifesting as features localized at  critical interfaces in the AFM topographs of Figs.~\ref{fig:fab_geometry}(g-i).  However, these byproducts adhere more strongly to the original SiO2 substrate than to the hBN flake used for pick-up.  This results in a self-cleaning process that enables the transfer of pristine microstructures into the middle layers of the heterostructure. Figs.~\ref{fig:fab_geometry}(j-l) show AFM topographs of the patterned graphite after pick-up by a hexagonal boron nitride flake--\textit{i.e.}, imaged with the graphite in the configuration shown in Fig. \ref{fig:fab_geometry}c.  Etch byproducts visible in the as-cut graphite (Fig.~\ref{fig:fab_geometry}g-i) are not transferred, leaving pristine nano-scale subtractive patterns.  Notably, in this process the areas closest to the critical regions are not exposed to additional fabrication residues, in contrast to graphite gates patterned by plasma-etching \cite{ronen_aharonov-bohm_2021}.

Remarkably, our process permits transfer of high-density patterns without degradation or tearing.  Three examples are shown in  Figs.~\ref{fig:fab_geometry}d-f, depicting a quantum point contact, quantum dot array, and Fabry-P\'erot interferometer, all meant to be operated in the quantum Hall regime.  We note that to preserve the alignment of fine lithographic features during pattern transfer, the patterned graphite flakes must remain contiguous: free-floating graphite pieces will typically slide during transfer leading to electrical shorts between nominally disconnected areas. In most cases, additional conventional subtractive processing is used to electrically disconnect different gate regions and make electrical contact to sample and gate layers.  However, certain desirable electrostatic geometries can be realized using complementary patterns in both the graphite top- and bottom- gates and aligning the two during transfer. An example of this strategy, used to construct a quantum Hall edge state Fabry-P\'erot interferometer, is shown in Figs.~\ref{fig:fab_geometry}f, i, and l, in which a patterned bottom gate is used to create an isolated, independently density-tunable region in a graphene device without requiring a free-floating gate.

While the process described above produces \textit{topographically} pristine gate geometries, topography alone is not a guarantee of electronic quality. In particular, without electrical characterization, we cannot exclude that the anodic oxidation itself produces unacceptably large local potentials on the graphite edge. To qualify our technique, then, we use a monolayer graphene quantum point contact (QPC) device operating in the quantum Hall regime.
A QPC is formed when a narrow constriction in a two-dimensional device restricts the number of quantized channels through which electrons can flow\cite{van_wees_quantized_1988}. In the quantum Hall regime, when the constriction is on the order of the magnetic length $\ell_B$, the transmission of chiral edge modes through the QPC varies sensitively with the width.
To form a QPC using our fabrication method, we start from the cross gate geometry of Fig. \ref{fig:fab_geometry}d, and perform additional lithographic processing (see Methods) to isolate the four quadrants of the graphite top gate producing four isolated gates we denote North (N), South (S), East (E) and West (W). Voltages applied to these gates and a global graphite bottom gate can be used to deplete the monolayer into the $\nu = 0$ gap in the $N$ and $S$ regions, forming a narrow constriction, with the filling factor in the $E$ and $W$ regions held constant. An optical micrograph of the completed device is shown in Fig.~\ref{fig:integer_qpc}a. 

To characterize the operation of our QPC we measure the four-terminal `diagonal conductance,' $G_{\mathrm{D}}$ (see Methods), which in the integer quantum Hall (IQH) regime gives a direct measure of the number of edge modes $N_{\text{qpc}}$ transmitted across the device such that $G_{\mathrm{D}} = N_{\text{qpc}} \frac{e^2}{h}$\cite{datta_electronic_1995, zimmermann_tunable_2017}.
Fig.~\ref{fig:integer_qpc}b shows $G_{\mathrm{D}}$ measured at $B=6$~T and T=300~mK as the gate voltages are adjusted to tune $N_{\text{qpc}}$ from 0 to 2.  In this measurement, we fix the value of $V_\mathrm{EW}+\alpha V_\mathrm{B}$ (where $\alpha=c_b/c_t$ is the ratio of the bottom- and top-gate capacitances to the monolayer, and $V_{\text{EW}}$ denotes a single voltage applied to the $E$ and $W$ gates -- see Fig.~\ref{fig:supp_determining_alpha}a-c). This fixes the  bulk filling factor in the $E$ and $W$ regions within the $\nu = -4$ plateau.  We also fix $V_\mathrm{NS}+\alpha V_\mathrm{B}$ to keep the $N$ and $S$ regions within the $\nu = 0$ plateau.  Tuning $V_{\text{EW}}-V_{\mathrm{B}}$ then controls the QPC width independent of the bulk filling factors in the four quadrants (see Fig.~\ref{fig:supp_integer_qpc} and Fig.~\ref{fig:supp_ext_integer_qpc} for additional info on QPC operation in the integer quantum Hall regime). 
As is evident in Fig.~\ref{fig:integer_qpc}b, we observe integer-quantized conductance plateaus separated by monotonic transitions; in particular, we do not observe non-monotonic conductance features characteristic of disorder-mediated or resonant tunneling effects, typical of quantum point contacts in both III-V semiconductors and graphene \cite{milliken_indications_1996, zimmermann_tunable_2017, ronen_aharonov-bohm_2021, radu_quasi-particle_2008, baer_interplay_2014}. 

\section{Chiral Luttinger liquid behavior at a fractional quantum point contact}

\begin{figure*}[ht]
    \includegraphics[width = 178mm]{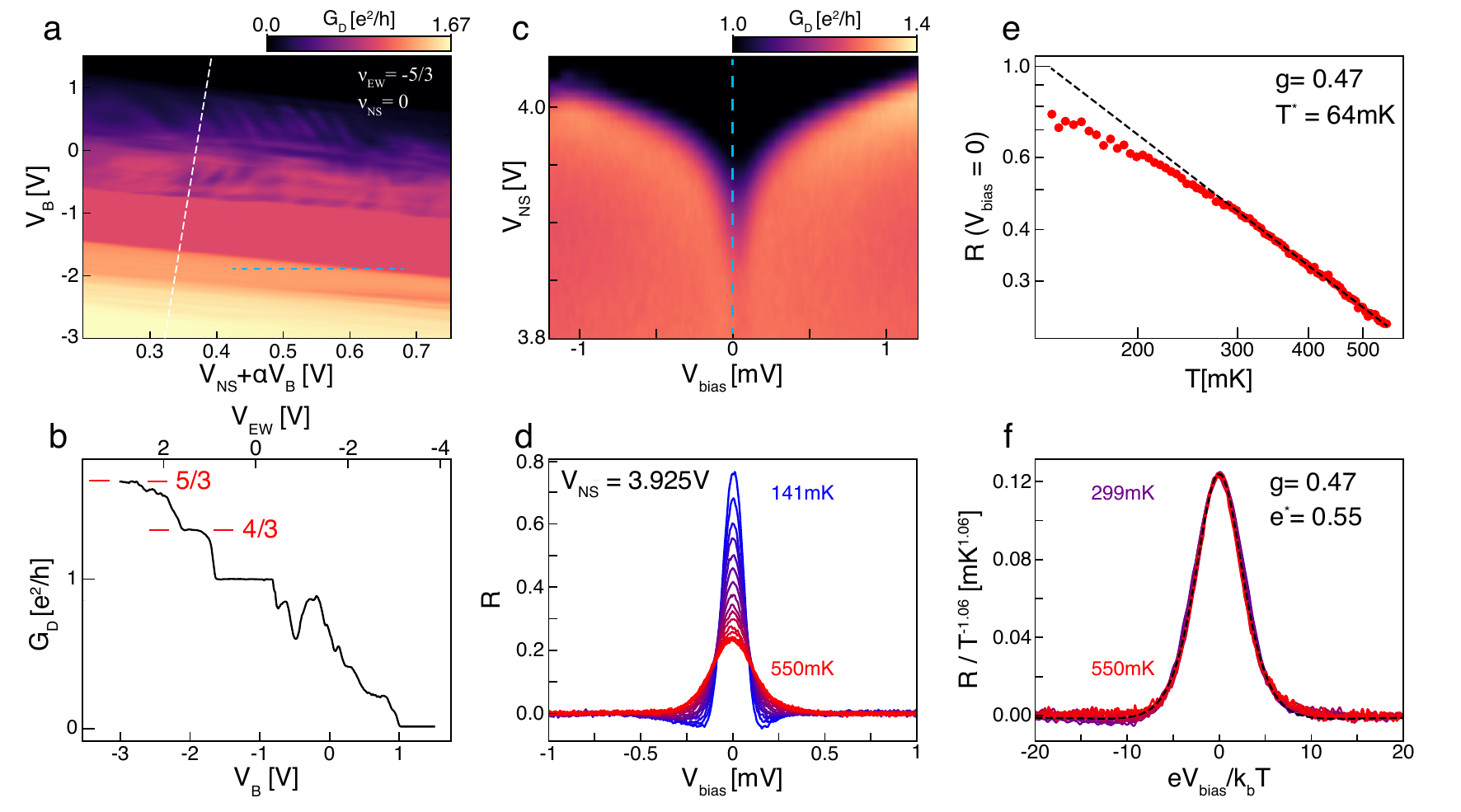}
    \caption{
    \textbf{Partitioning of fractional quantum Hall edges and quasi-particle tunneling.} 
    \textbf{(a)} $G_{\mathrm{D}}$ map at fixed $\nu_{EW} = -\frac{5}{3}$.  As $V_{\mathrm{B}}$ is adjusted along the y-axis, $V_{\text{EW}}$ is swept in tandem to keep $V_{\text{EW}} + \alpha V_{\mathrm{B}} \propto \nu_{EW}$ constant while allowing $V_{\mathrm{B}}-V_{\text{EW}}$ to vary, adjusting the saddle point electrostatics.  On the x-axis, the range of $V_{\text{NS}}$ is varied as $V_{\mathrm{B}}$ changes in order to keep the range $V_{\text{NS}} + \alpha V_{\mathrm{B}}$ within the $\nu = 0$ plateau.
    \textbf{(b)} $G_{\mathrm{D}}$ along the dashed white contour in panel a.  Two fractional plateaus are highlighted at $4/3$ and $5/3$ resepctively. 
    \textbf{(c)} Bias dependence of the $G_{\mathrm{D}} = \frac{4}{3}$ plateau, along the dashed blue line shown in panel a. The tunneling conductance is suppressed at zero bias well into the plateau, a distinct signature of chiral Luttinger liquid behavior.  
    \textbf{(d)} Extracted back-scattered conductance, $R$, versus scaled DC bias, $V_{\text{bias}}$, at a fixed $V_{\text{NS}} = 3.925V$ for the temperatures $T = $ \SI{141}{mK}, \SI{169}{mK}, \SI{199}{mK}, \SI{229}{mK}, \SI{259}{mK}, \SI{289}{mK}, \SI{320}{mK}, \SI{350}{mK}, \SI{379}{mK}, \SI{410}{mK}, \SI{440}{mK}, \SI{470}{mK}, \SI{499}{mK}, \SI{530}{mK}, \SI{534}{mK}, \SI{540}{mK}, \SI{544}{mK}, \SI{550}{mK}. \textbf{(e)} Back-scattered conductance, $R$, at zero DC bias versus temperature.  Power-law behavior onsets above \SI{300}{mK}; we find $R \propto T^{-1.057 \pm 0.008}$, yielding a Luttinger parameter of $g = 0.47 \pm 0.004$.  The cross-over energy scale between weak electron tunneling and weak quasi-particle back-scattering is estimated to be $T^* = 64 \pm \SI{3}{mK}$ (notably power-law behavior with negligible contributions from electron tunneling for quasi-particle back scattering is not expected unless $T >> T^*$).  \textbf{(f)} Scaling collapse of DC bias dependent tunneling curves for different temperatures above power-law onset.  Here $R / T^{-1.06}$ is plotted against $e V_{\text{bias}} / k_b T$ for the temperatures $T = \SI{299}{mK}, \SI{320}{mK}, \SI{340}{mK}, \SI{360}{mK}, \SI{379}{mK}, \SI{399}{mK}, \SI{420}{mK}, \SI{440}{mK}, \SI{460}{mK}, \SI{480}{mK}, \SI{499}{mK}, \SI{520}{mK}, \SI{540}{mK}$.  The black-dashed is a fit to the expected bias dependence for a chiral Luttinger liquid with g = 0.47 (see Eq.~\eqref{eqn:digamma}) yielding an effective quasi-particle charge of $e^* = (0.55 \pm 0.001) |e| $. 
    }
    \label{fig:fractional_qpc}
\end{figure*}

\begin{figure}[ht]
    \centering
    \includegraphics[width = 180mm]{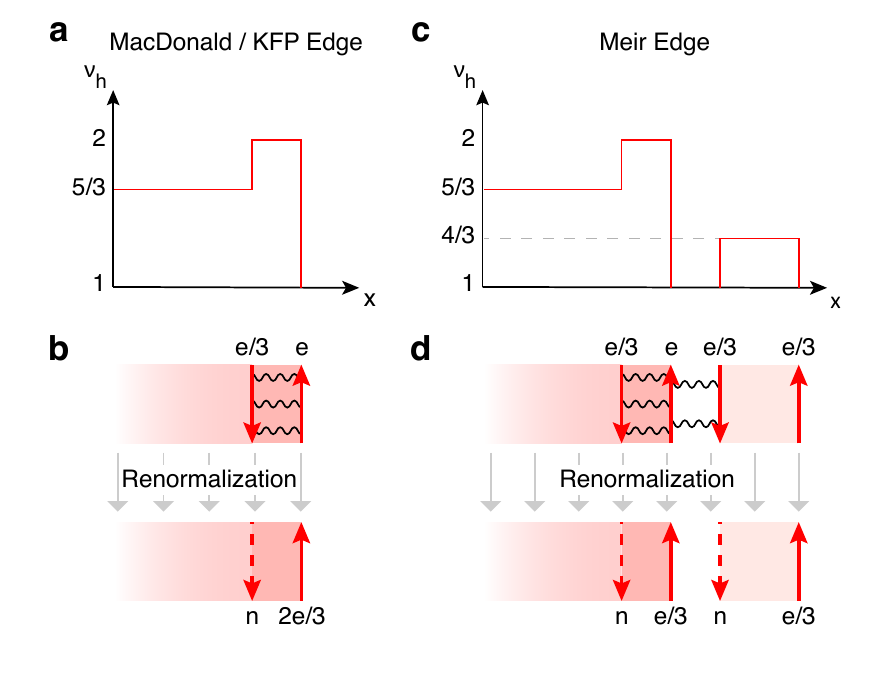}
    \caption{\textbf{Possible Edge Structures in $\nu = -5/3$} 
    \textbf{(a)} Hole-conjugate FQH states such as $|\nu| = 2/3, 5/3$ states can be modeled by a Laughlin-like FQH state of holes within a bulk integer quantum Hall state, leading to a small strip of increased filling factor around the edge of the sample. This is shown schematically in panels a and c by plotting the filling factor of holes $\nu_{h} \equiv -\nu$ at the boundary between a $\nu = -5/3$ and $\nu = -1$ state, where the relevant fractional edges measured in the experiment occur. The \textit{MacDonald model} \cite{johnson_composite_1991} of the resulting edge structure posits a downstream integer mode at the outermost edge of the sample, as well as an upstream (counter-propagating) fractional mode.
    \textbf{(b)} In real experiments, the two counter-propagating charged modes are rarely observed, but rather mix through the presence of inter-edge interactions, yielding a single effective charge-2e/3 mode propagating downstream, as well as an upstream charge-neutral mode, as explained by the \textit{Kane-Fisher-Polchinski} model \cite{kane_randomness_1994}.
    \textbf{(c)} A sufficiently soft confining potential may make it energetically favorable to redistribute the charge in the system and create an additional strip of density $\nu_{h}=4/3$, introducing a set of two additional counter-propagating fractional edge modes: the \textit{Meir model}. \cite{meir_composite_1994}
    \textbf{(d)} In a real system, where these modes can also mix, the resulting mode structure may contain two downstream fractional-conductance modes as well as two upstream neutral modes. This scenario is consistent with the observation of multiple fractional conductance steps within the $\nu = -5/3$ state.
    \label{fig:supp_edge_structure}
    }
\end{figure}

An even more stringent characterization of disorder at the QPC can be obtained in the  fractional quantum Hall (FQH) regime. Fig. \ref{fig:fractional_qpc}a shows $G_{\mathrm{D}}$ plotted as a function of $V_{\mathrm{B}}$ and  $V_{\text{NS}} + \alpha V_{\mathrm{B}}$ at B=13T and T=300mK. Throughout this range, the filling factor of the $E$ and $W$ region is fixed at $\nu=-5/3$, and that of the $N$ and $S$ regions is fixed within the $\nu=0$  plateau.  Data taken along the white dashed line in  Fig.~\ref{fig:fractional_qpc}a is plotted in Fig.~\ref{fig:fractional_qpc}b, and shows two distinct quantized plateaus at $G_{\mathrm{D}} = 4/3$ and $5/3$ , in addition to the integer plateau at $G_{\mathrm{D}} = 1$.  This observation suggests the $\nu = -5/3$ state hosts two edge modes each with conductance $-\frac{e^2}{3h}$.  Notably, this is not consistent with theoretical expectations for a sharp confining potential\cite{johnson_composite_1991, kane_randomness_1994}. In that limit, a pair of counter propagating edge modes is expected where one carries a conductance of $-\frac{e^2}{h}$ and the other carries a conductance of $\frac{e^2}{3h}$ (see Fig.~\ref{fig:supp_edge_structure}a); evidence for this structure was observed recently in carefully designed III-V heterostructures\cite{nakamura_half-integer_2022}. Our observation is instead consistent with a picture of the 5/3 edge which includes the effects of edge state reconstruction resulting from a soft confinement potential defining the boundary of the topological bulk\cite{sabo_edge_2017}.  
Theoretically, softly confined edges of hole-conjugate fractional quantum Hall states are expected\cite{chamon_sharp_1994, jianhui_wang_edge_2013, meir_composite_1994} to host neutral modes along with the observed fractionally quantized charged modes  (see Fig.~\ref{fig:supp_edge_structure}b). 
While these neutral modes do not directly couple to an applied electric field, they may affect thermal transport or renormalize the tunneling spectra of nearby charged modes \cite{kane_edge-state_1996}.

The edge modes at the boundary of a fractional quantum Hall state are expected to behave like chiral Luttinger liquids\cite{wen_chiral_1990, kane_edge-state_1996}. In this state of matter, the quasi-particle wave functions are  orthogonal to the electrons from  which they are microscopically constructed.  
This `orthogonality catastrophe' results in a soft gap in the electron tunneling density of states which vanishes like $\rho \propto (E - E_F)^{1/g -1}$, where the constant $g$ is known as the Luttinger parameter \cite{luttinger_exactly_1963}.  
A remarkable consequence of this fact is that even arbitrarily weak barriers between edge modes will suppress tunneling at sufficiently low temperature and bias voltage. 

\begin{figure}[ht]
    \centering
    \includegraphics[width = 180mm]{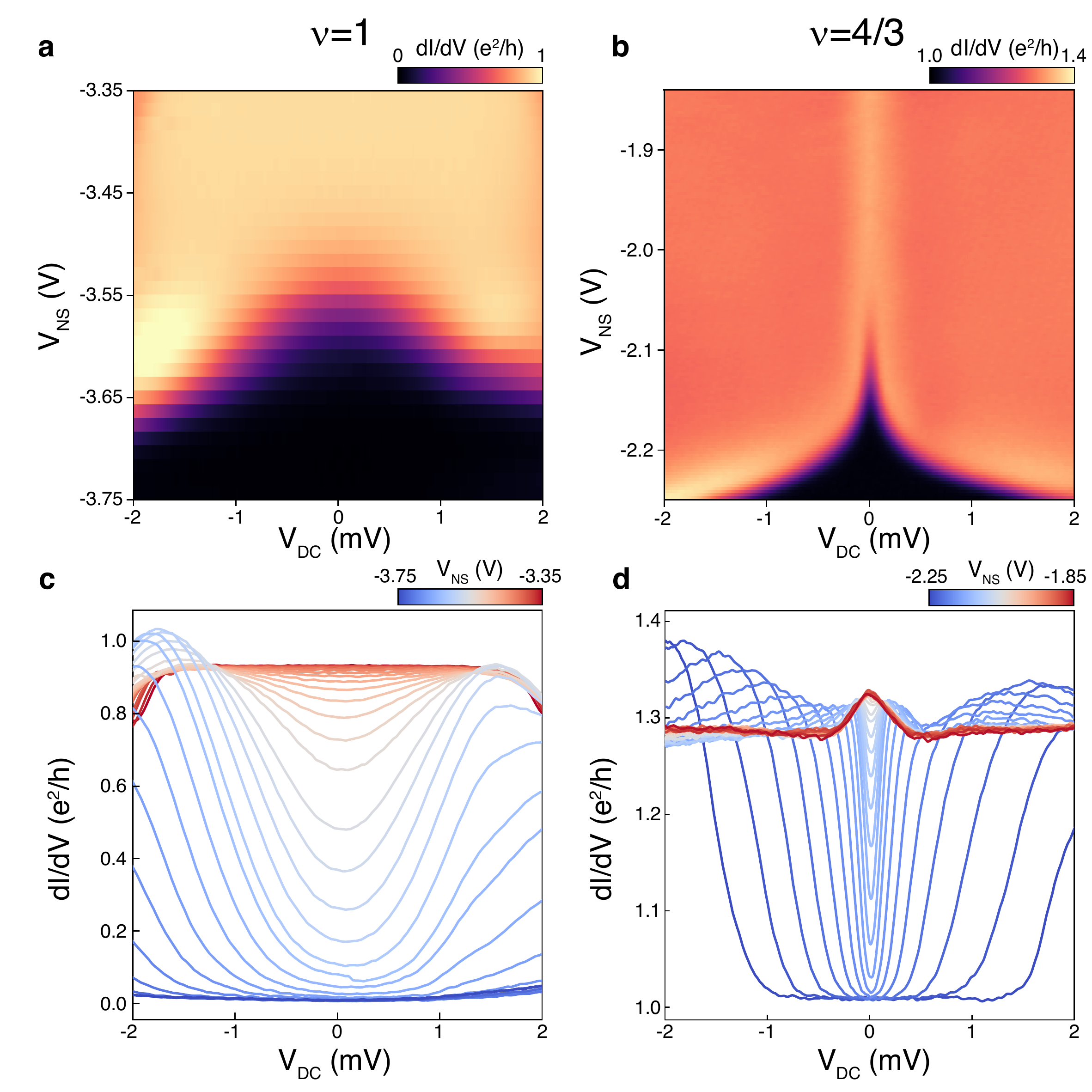}
    \caption{\textbf{Tunneling conductance in an integer vs. fractional edge} 
    \textbf{(a)} Plot of the tunneling conductance across an integer conductance step, in the $\nu$=1 state. In the fully reflecting and fully transmitting limits, the conductance is constant for $V_{DC}$ less than about |1~mV|, and smoothly varies as the edge is transmitted.
    \textbf{(b)} For a fractional edge state, the conductance remains highly suppressed even when the edge state is partially transmitted, with a sharply nonlinear $dI/dV$ near $V_{DC}$=0. Even when the edge state is fully transmitted, and $dI/dV (V_{DC} = 0) = 4/3$, the tunneling conductance remains nonlinear.
    \textbf{(c)} and \textbf{(d)} present linecuts of the data in (a) and (b) respectively for comparison.
    \label{fig:supp_integer_vs_fraction}
    }
\end{figure}

Fig.~\ref{fig:fractional_qpc}c shows $G_{\mathrm{D}}$ as a function of source-drain voltage $V_{\text{bias}}$  and $V_{\text{NS}}$, corresponding to the dashed blue trajectory in Fig.~\ref{fig:fractional_qpc}a.  The data  shows a single sharp zero-bias  conductance dip  throughout the transmission plateau at $G_{\mathrm{D}}  \approx \frac{4}{3} \frac{e^2}{h}$--in contrast with the weak bias dependence observed for an integer edges (see Fig.~\ref{fig:supp_integer_vs_fraction}). In Fig.~\ref{fig:fractional_qpc}c, $V_{\text{NS}}$ tunes the electrostatic confinement at the QPC and thus the height of the tunnel barrier separating the two outermost edge modes incident upon the junction.  As expected, a decrease in $V_{\text{NS}}$ lowers the potential barrier at the QPC and enhances transmission. 
The lack of resonant structure in Fig.~\ref{fig:fractional_qpc}c, features observed in previous experiments\cite{milliken_indications_1996, radu_quasi-particle_2008, baer_interplay_2014,zimmermann_tunable_2017}, even at high transmission suggests that scattering between edge modes occurs at a single, gate-controlled saddle point. 

To compare our data with expectations from the chiral Luttinger liquid theory\cite{wen_edge_1991,kane_edge-state_1996}, Fig.~\ref{fig:fractional_qpc}d shows the reflection coefficient, $R$, which measures the probability of backscattering the outermost fractional edge mode (see also Fig. \ref{fig:supp_tunneling_analysis} for a detailed analysis, which follows Refs. \onlinecite{wen_edge_1991,roddaro_nonlinear_2003,radu_quasi-particle_2008,baer_experimental_2014}).  
For two fractional quantum Hall edges which are weakly tunnel coupled via the exchange of quasi-particles, the reflection coefficient $R$, as a function of $V_{\text{bias}}$, may be computed perturbatively \cite{wen_edge_1991}. Evidently, even for high transmission, voltage bias alone can tune $R$ from over 75\% to nearly zero, demonstrating that the incident edge modes become decoupled a low energies -- a manifestation of the `orthogonality catastrophe' -- a hallmark of chiral Luttinger liquid physics. 

At zero bias, when quasi-particle backscattering is weak (corresponding to small $R$), $R$ is expected to follow a simple power law with respect to temperature, $R\propto T^{2g-2}$.  We plot the zero-bias $R$ in Fig.~\ref{fig:fractional_qpc}e.  For $T$ between 300-550mK---corresponding to $R\lesssim 0.4$---the reflection coefficient is well fit by a power law with $g=0.47$.  In this weak-backscattering regime, the dependence of $R$ on $V_{\text{bias}}$ and $T$ is theoretically expected\cite{wen_edge_1991} to follow a scaling form, 
\begin{multline}
    \frac{R(x)}{T^{2g - 2}} = A \beta (g + i \frac{e^* x}{2 \pi}, g - i \frac{e^* x}{2 \pi}) \times \\
    \bigg[\pi \cosh(\frac{e^* x}{2}) + 2 \sinh (\frac{e^* x}{2}) \text{Im} [\Psi (g + i\frac{e^* x}{2 \pi})]\bigg]
    \label{eqn:digamma}
\end{multline}
where $x=\frac{eV_{\text{bias}}}{k_bT}$ is the scaled voltage bias, $\beta$ and $\Psi$ denote the corresponding Euler integrals, $e^*$ is the effective quasi-particle charge in units of the electron charge, and A is a constant related to the height of the tunnel barrier.  

It follows from this equation that if $R(x)$ is scaled appropriately by the temperature, tunneling spectra taken in the temperature range where the zero-bias power law applies should follow a universal curve, with $e^*$ providing a single additional adjustable parameter. Fig.~\ref{fig:fractional_qpc}f shows $R / T^{-1.06}$ versus $e V_{\text{bias}} / k_b T$ for several temperatures in the range of $\SI{300}{mK}$ to $\SI{550}{mK}$.  The curves are observed to collapse onto each other fixing $g = 0.47$. The collapsed curves are then averaged and fit to Eq.~\eqref{eqn:digamma} extracting the quasi-particle charge as $e^* = 0.55 \pm 0.001$.

The observed scaling collapse provides evidence of tunneling between chiral Luttinger liquids at a single point within the QPC.  However, the measured values of $g = 0.47$ and $e^* = 0.55$ are at odds with predictions that that the boundary of a fractional quantum Hall phase should host a chiral Luttinger liquid with a quantized Luttinger parameter and quasi-particle charge, namely $g=1/3$ and $e^*=1/3$\cite{wen_chiral_1990}. We attribute this discrepancy to the presence of neutral modes at the $\nu = -5/3$ boundary, consistent with an electrostatically shallow edge permitting significant Coulomb-induced reconstruction effects\cite{kane_edge-state_1996, meir_composite_1994} (See Fig.~\ref{fig:supp_edge_structure}b).  While a universal regime is always expected to exist, for the electrostatic configuration explored here, this regime most likely occurs significantly below $\SI{300}{mK}$, outside of the weak quasi-particle backscattering limit.  Notably, past experiments on a variety of fractional Hall states in III-V semiconductor systems\cite{chang_observation_1996, chang_plateau_2001, milliken_indications_1996, radu_quasi-particle_2008,chang_observation_1996, chang_plateau_2001} have also shown significant deviations from the expected universal behavior. The deviation from universality in all of these experiments, including the one presented here, is likely traceable to the complex edge structure arising from the shallow edge confinement typical of gate-defined tunnel barriers. 

\begin{figure}[ht!]
    \includegraphics[width = 88mm]{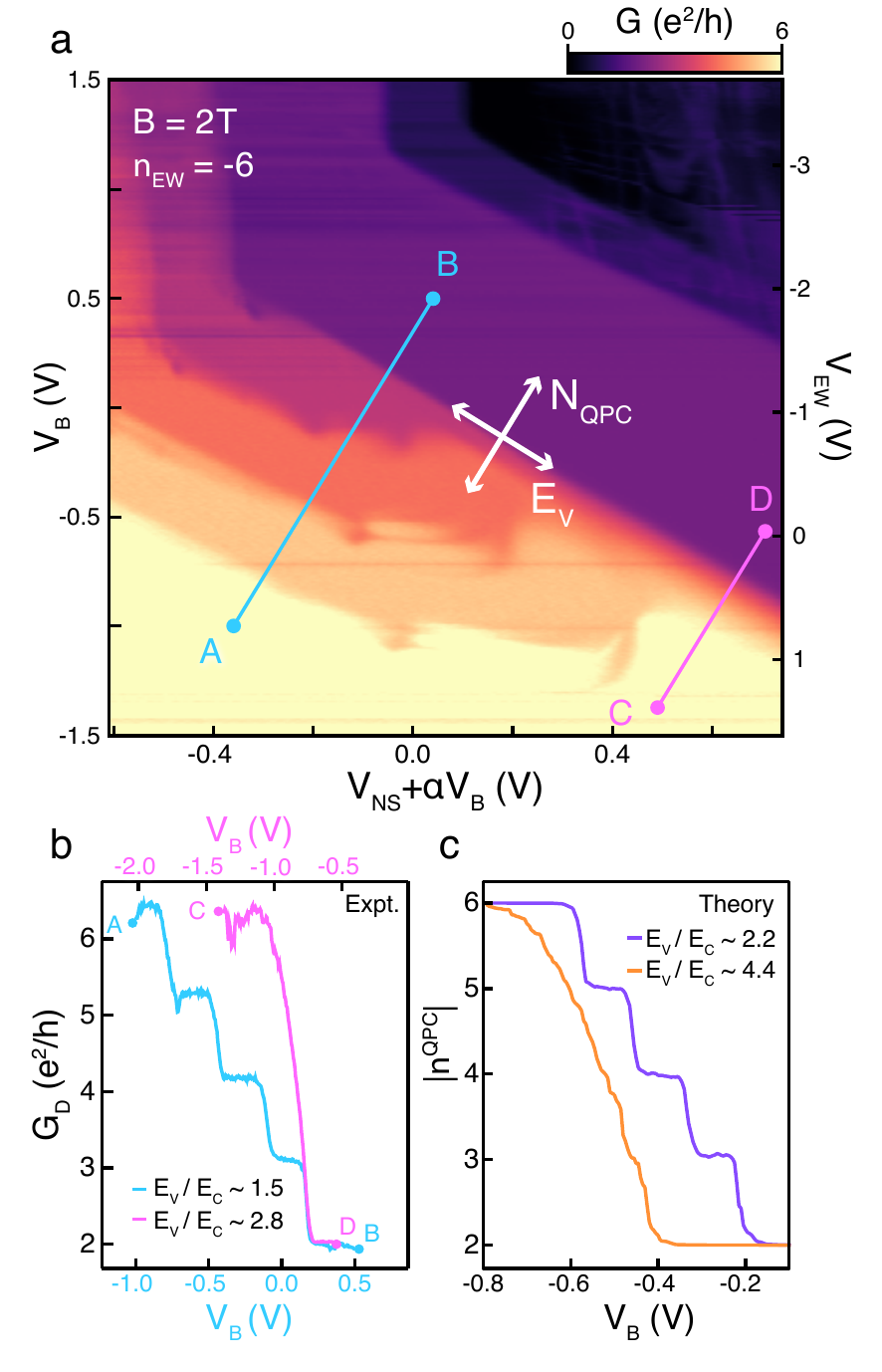}
    \caption{
    \textbf{Tuning edge sharpness via electrostatic gating.} 
    \textbf{(a)}. 
    $G_{\mathrm{D}}$ map at fixed $\nu_{EW} = -2$ at B = 2T.  As in Fig.~\ref{fig:fractional_qpc}a, the gates are swept to maintain constant filling in the E, W regions while varying $N_{QPC}$ and the sharpness of the confining potential.
    \textbf{(b)} Data of panel a along the marked trajectories, showing a qualitative change from individual $1\frac{e^2}{h}$ steps to a single continuous transition from $G_{\mathrm{D}} = 2\frac{e^2}{h}$ to $G_{\mathrm{D}} = 6\frac{e^2}{h}$
    \textbf{(c)} Thomas-Fermi simulation of the filling factor at the center of the QPC. 
    }
    \label{fig:2T_electrostatic}
\end{figure}

This deficiency may be addressed in our geometry using \textit{in situ} control of the confinement potential. 
The softness of the confinement potential is quantified by the ratio of $E_V \equiv e\frac{\partial \phi_{ext}}{\partial x}\ell_B$ (here $x$ is the coordinate perpendicular to the confined edge, and $\ell_B$ is the magnetic length) and the Coulomb energy $E_C \equiv \frac{e^2}{4\pi \epsilon\ell_B}$. In our geometry, independent control of the voltages applied to the N/S, E/W, and bottom gates is equivalent to independent control of the filling factor in the E/W regions, the number of transmitted modes $N_{\text{qpc}}$, and $E_V$, allowing us to explore the effects of the confinement energy within the QPC independently of the barrier height.  

To illustrate this, Figure \ref{fig:2T_electrostatic}a shows a map of $G_{\mathrm{D}}$  measured at $B = 2$~T with fixed $\nu = -6$ in the E/W regions. 
In this plot, moving along a diagonal from lower left to upper right reduces $N_{\text{qpc}}$, while moving along a diagonal from upper left to lower right increases $E_V$. 
We observe a qualitative change in the nature of the transition between $G_{\mathrm{D}}=2\frac{e^2}{h}$ to $G_{\mathrm{D}}=6\frac{e^2}{h}$ as a function of $E_V$.  
For small $E_V$---\textit{i.e.} soft confinement---we observe distinct conductance plateaus corresponding to each integer quantum Hall state between $\nu = -2$ and $\nu = -6$, 
indicating that the four distinct edge modes corresponding to the $\nu = 3$ through $\nu = 6$ states can be individually partitioned.
For large $E_V$, in contrast, 
the steps merge into one continuous jump from $G_{\mathrm{D}} = 2\frac{e^2}{h}$ to $G_{\mathrm{D}} = 6\frac{e^2}{h}$, with no discernible intermediate plateaus at partial transmission (see Fig.~\ref{fig:2T_electrostatic}b).

This behavior can be understood in the context of the competition between $E_V$ and $E_C$ at the QPC.  In the regime of Fig. \ref{fig:2T_electrostatic}, in the bulk, the four-fold spin and valley degeneracy of the Landau levels is lifted due to quantum Hall ferromagnetism\cite{dean_fractional_2020}.  In the presence of a soft confining potential, the electron system can lower its energy by forming islands of incompressible fluid at integer filling factor within the QPC saddle point, due to the emergence of $E_C$-scale integer quantum Hall ferromagnetic gaps. However, as $E_V$ is increased it is no longer energetically favorable to form such incompressible islands within the QPC, and instead a smooth change in electron density at the QPC as function of transmission is preferred.  Finite element electrostatics calculations show $E_V/E_C$ is approximately $1.5$ and $2.8$ for the trajectories marked AB and CD in Fig.~\ref{fig:2T_electrostatic}a, respectively, where we take $E_V$ from the maximum value of the potential slope at the boundary between the N and W regions.   

To assess the plausibility of this explanation, we perform simulations of the electron density in our device within the Thomas-Fermi approximation (see Supplementary Information Section 5 for details of the simulation). 
These simulations take as inputs the device electrostatics as well as a phenomenological model for the graphene chemical potential, which influences the quantum capacitance. 
For $E_V/E_C = 2.2$, the simulated density in the vicinity of the QPC exhibits a ``wedding cake'' profile, with well defined strips of integer filling, while for larger $E_V/E_C = 4.4$ the density drops sharply in a single step across the quantum point contact (see Fig. \ref{fig:tf_model_summary_2}). 
% The discrepancy of the critical $E_V/E_C$ between the experiment and the simulation could originate from the screening from virtual inter-LL  transitions that had been neglected in the simulation.
% The corresponding charge density profiles of these two cases are shown in Fig.~\ref{fig:tf_model_summary_2}b-c. 
As a heuristic proxy for $G_D$, we plot the simulated filling factor at the center of the QPC, $|\nu^{\textrm{QPC}}|$, in Fig. \ref{fig:2T_electrostatic}c. 
The soft confinement regime shows a series of integer steps, while sharp confinement does not. This indicates that only in the case of soft confinement do the edge-states remain well separated at the QPC, allowing their transmission to be partioned individually as observed in Fig.~\ref{fig:2T_electrostatic}a. 
% Moreover, it highlights that in our device we may adjust $E_V$ to a point where the density profile is primarily defined by the external potential, rather than many-body effects.

As discussed previously, rearrangement of the electron density also causes significant effects on the edge in the fractional quantum Hall regime.  These effects, referred to generally as ``edge-reconstruction,'' can nucleate additional edge modes at soft potential boundaries with quantitative effects on tunneling behavior.   We estimate $E_V/E_C \sim 1.3$ for the fractional quantum Hall tunneling experiment of Fig. \ref{fig:fractional_qpc}, squarely in the regime where edge reconstruction effects are expected. Adjusting $E_V$ such that the effects of $E_C$ at the edge are negligible should allow exploration of universal edge mode physics in both tunneling and interferometry experiments.

\section{Conclusions}

In summary, we have shown that AFM-based local oxidation lithography\cite{li_electrode-free_2018}---applied to  graphite and coupled with a low-strain dry transfer procedure---allows for nanoscale features to be fabricated in the gating  layers of a van der Waals heterostructure, while keeping the disorder level sufficiently small to resolve the subtle physics of tunneling between chiral Luttinger liquids at a single point. Control over $E_V$, in addition to $N_{\text{qpc}}$ and the bulk filling $\nu_{EW}$, provides a new knob to turn which likely will enable operation of quantum Hall mesoscopic devices in regimes where universal physics is expected and the behavior is significantly simpler to interpret than in previous measurements \cite{chang_observation_1996, chang_plateau_2001}. We expect these techniques may be immediately applied to direct probes of topological order, as quantum point contacts constitute the essential building block for measuring the Luttinger liquid parameter \cite{radu_quasi-particle_2008}, quantized thermal Hall response\cite{banerjee_observation_2018,banerjee_observed_2017}, and interferometry\cite{nakamura_direct_2020}.  More generally, however, the techniques we describe enable a new level of complexity in device design, allowing the equivalent of monolithic integrated circuits built from gate-tunable correlated phases in van der Waals heterostructures. 

%TC:ignore

\section{Acknowledgements}
The authors thank J. Folk and A. Potts for comments on the manuscript, and C.R. Dean and J. Swan for advice on humidity stabilization.  We additionally acknowledge Youngjoon Choi for providing AFM topographs of quantum dot arrays.
Work at UCSB was primarily supported by the Air Force Office of Scientific Research under award FA9550-20-1-0208 and by the Gordon and Betty Moore Foundation EPIQS program under award GBMF9471. 
L.C. and N.S. received additional support from the Army Research Office under award W911NF20-1-0082. 
T.W. and M.Z. were supported by the Director, Office of Science, Office of Basic Energy Sciences, Materials Sciences and Engineering Division of the U.S. Department of Energy under contract no. DE-AC02-05-CH11231 (van der Waals heterostructures program, KCWF16).
K.K. was supported by the U.S. Department of Energy, Office of Science, National Quantum Information Science Research Centers, Quantum Science Center. 
C.R. was supported by the National Science Foundation through Enabling Quantum Leap: Convergent Accelerated Discovery Foundries for Quantum Materials Science, Engineering and Information (Q-AMASE-i) award number DMR-1906325. 
K.W. and T.T. acknowledge support from JSPS KAKENHI (Grant Numbers 19H05790, 20H00354 and 21H05233).

\section{Author Contributions}
L.A.C and A.F.Y conceived of the experiment.  L.A.C and N.S. fabricated the devices and performed the measurements.  L.A.C, N.S, M.Z., T.W., K.K., and A.F.Y analyzed the data and wrote the paper.  M.Z. and S.V. proposed the simulation methodology.  T.W., K.K., and C.R. performed numerical calculations based on the proposed theoretical modeling.  T.T. and K.W. synthesized the hexagonal boron nitride crystals

\section{Competing Interests}
The authors declare no competing interests.

\section{Tables}
\section{Figure Legends/Captions}

\newpage
\normalem
\section{References}
\bibliographystyle{custom}
\bibliography{references, recon_bib}
%\putbib[references]
%\end{bibunit}

\section{Methods}
%\begin{bibunit}[custom]

\subsection{van der Waals Heterostructure Assembly}
Graphene and hBN were mechanically exfoliated from bulk crystals using a combination of thermal release tape and scotch magic tape.  The initial mother-tape is prepared using 3M Scotch-brand magic tape for graphite or 3M Scotch-brand greener magic tape for hBN.  For hBN, thermal release tape is adhered to the mother-tape to generate a daughter tape.  For graphite, the daughter-tape is made with 3M Scotch-brand magic tape.  The daughter-tape is removed from the mother-tape, cleaving the bulk crystals along the c-axis, then transferred onto a 1cm x 1cm doped Si chip with 285nm of thermally-grown SiO\textsubscript{2} on the surface.  For graphite, the substrate is heated to $110^{\circ}$ C for 60 seconds, before removing the tape quickly to reduce glue residue remaining on the SiO\textsubscript{2} surface.  For hBN, the daughter-tape is transferred onto the SiO\textsubscript{2} surface at room temperature and is removed from the SiO\textsubscript{2} slowly.  The Si/SiO\textsubscript{2} substrates are cleaned by a standard solvent process: the chip is cleaned in acetone for 5 minutes in a high power ultrasonic bath, followed by an IPA wash, and finished with an N2 blow-dry. Additionally, for graphite the SiO\textsubscript{2} surface is treated in O\textsubscript{2} plasma at 100W and 300mTorr for 60 seconds in order to promote adhesion of large multilayers.  The resulting exfoliated crystals are then characterized by optical microscopy.

AFM-LAO was performed on a Bruker Dimension Icon AFM.  Exfoliated graphite flakes, prepared as described above, are loaded into the AFM.  The humidity is controlled using a bang-bang style humidity controller.  
The plant is formed by a beaker filled with 250mL of deionized water placed on a hot plate at $120^{\circ}$ C.  
Once the humidity sensor measures higher than 50\% RH, the hot-plate is switched off.  We pattern sub 100 nm  crosses into a 3 nm  graphite flake to form the top gate of the quantum point contact.  This is accomplished using a Pt/Ir coated Arrow-NcPt AFM probe from nanoandmore.  A topographical map of the graphite flake is obtained and the cross pattern is placed in an area with no visible defects.  The lithography is performed in Bruker's Nanoman software package which allows for precise control of the direction, speed, and deflection of the conducting AFM probe.  Graphite can be etched with AFM-LAO under a variety of conditions, however we have found that lithography performed in contact mode with an 18V peak-to-peak excitation at 150kHz provided the smallest line-widths achievable in our system.  For new AFM probes, typical line-widths are on the order of 60-70~nm .  This leads to a QPC critical dimension (tip-to-tip distance) between 90 and 100~nm .  However, due most likely to hydrocarbon build up or natural wear on the AFM probe, the cut-width broadens to 100~nm  after $\sim$150 $\mu$m of cutting.  

Before assembling the van der Waals heterostructure we fabricate a `transfer-slide:' a PDMS stamp adhered to a glass slide with a polycarbonate laminate transferred on top used to `pick-up' the first layer of hBN.  Initially, 8g of Sylgard 184 PDMS is mixed in a 10:1 ratio by weight with a curing agent and poured into a standard 100~mm plastic petri dish.  The PDMS is left to cure at room temperature for 24 hours in a vacuum chamber in order to remove any bubbles that formed during mixing.  Additionally, another 3g of Sylgard 184 PDMS solution is mixed with a curing agent and left to cure partially at room temperature under vacuum for 2 hours.  A PDMS cylinder is cut out using a 2~mm hole punch from the 8g PDMS batch and is then adhered to a glass slide using the partially cured PDMS.  An additional droplet of partially cured PDMS is pipetted onto the cylinder in order to form a dome.  Slides are left to finish curing for another 24 hours at room temperature.  The resulting slides are then inspected under an optical microscope for dirt particulates.

The polycarbonate laminate is made from a 13.3\% wt/vol ratio of Bisphenol-A polycarbonate (PC) in cyclopentanone.  Unaided, cyclopentanone will only dissolve a 4\% wt/vol solution of PC, however by using the QSonica 500 ultrasonic wand, solutions up to 15\% wt/vol can be made.  This recipe was derived from previous work by Abbas \cite{abbas_nanofabrication_2013}.  Due to its significantly lower vapor pressure than chloroform, PC dissolved in cyclopentanone can be effectively spun onto a Si substrate.  Thin films of PC are prepared by spinning the mixture onto a Si substrate at 1400 rpm for 60 seconds.  The thin film is then subsequently transferred onto a domed PDMS stamp.  The transfer slide is heated between $160^{\circ}$C to $180^{\circ}$C for 5 minutes in order to increase the adhesion between the PC film and the PDMS stamp as well as to remove any air bubbles that may have formed during transfer of the film.

We begin the assembly of the van der Waals heterostructure by picking up an hBN flake larger than 100um in both width and length.  Ensuring the first hBN has the largest area aids in assembling the remaining layers since subsequent crystals will adhere entirely to another van der Waals material.  The transfer slide is engaged at a $0^{\circ}$ tilt angle, with the substrate heated to $70^{\circ}$C, and with the touch-down point 100-150um away from the center of the first hBN layer.  Once the transfer slide is brought into near contact the substrate is heated to $105^{\circ}$C.  This causes the transfer slide to fully engage and laminate over the first hBN layer.  The substrate is then cooled naturally back down to $70^{\circ}$ which retracts the transfer slide and removes the target hBN flake from the SiO\textsubscript{2}.  Next, the AFM-LAO etched graphite top gate is picked up while entirely encapsulated by the large initial hBN layer.  There are a number of advantages to using a domed PDMS transfer slide, one of which is that the engage point is in the center of the PDMS stamp, as opposed to the edge, which has fewer dust particulates that can interfere with assembly.  Moreover, domed PDMS transfer slides are known to reduce strain during vDH assembly \cite{saito_independent_2020}; it turns out this technique is critical for picking up AFM-LAO nano-structures without inducing tears or folds.

The PC laminate is extremely uniform as a result of being spun onto a Si substrate.  This also minimizes a common issue in assembling van der Waals heterostructures where the polycarbonate exhibits stochastic adherence to the SiO\textsubscript{2} substrate.  This often causes a `jerk' like motion during stacking which easily can cause graphite nano-structures to tear.  The remaining layers are assembled in the same manner; the full resulting device stack can be seen in Fig.~\ref{fig:fab_geometry}c. The stack is then deposited onto a doped silicon substrate with 285nm of thermally grown oxide, which forms the basis for a global bottom gate used to dope the graphene contacts.  This is accomplished by engaging the transfer slide and heating the substrate to $180^{\circ}$C (above the glass transition temperature of PC) to detach the PC from the PDMS stamp.  The laminate is then dissolved in chloroform for a minimum of 30 min.  Afterwards, the sample is rinsed in acetone and IPA, and then blown-dry with N2.

\subsection{Device Post Processing}
Heterostructures are post-processed using standard electron beam lithography, vacuum deposition, and dry-etching processes.  A device-defining plasma etch is used to separate the nanotextured graphite into four quadrants that we label North (N), South (S), East (E), and West (W): this is accomplished by inductively coupled plasma (ICP) etching in 40 sccm of CHF\textsubscript{3} and 4 sccm of O\textsubscript{2}.  This etch also separates the graphene contacts C1-8 such that any two contacts are only connected through the dual-gated region.  Additionally, the conducting Si substrate is used to dope each contact and prevent the formation of p-n junctions at the boundary between the contacts and the dual-gated region.  Finally, several trenches are etched across the boundary to the dual-gated region to introduce local scattering sites that improve equilibration between the contacts and quantum Hall edge modes in the device region. 

The etch mask is patterned by lifting off 40nm of Al using a polymethyl methacrylate (PMMA) A4 495K / A2 950K bilayer resist.  The PMMA is exposed using a 30kV electron-beam in an FEI SEM at 4004uC/cm\textsuperscript{2} and developed in a mixture of DI:IPA 1:3 kept at $10^{\circ}$C.  Al is deposited at 0.3A/s and the lift off is done for over 12 hours in N-Methyl-2-Pyrrolidone (NMP).  Post etching, the Al is dissolved in AZMIF300 photoresist developer which contains $<3$\% by weight of Tetramethylammonium hydroxide (TMAH).  Edge contacts are deposited onto the exposed graphene contacts and graphite gates using the same bilayer PMMA mask.  Before vacuum deposition, a brief contact cleaning etch in an ICP with 40 sccm of CHF\textsubscript{3} and 9 sccm of O\textsubscript{2} is performed.  Subsequently, a metal stack of Cr/Pd/Au 3/15/185 nm  is deposited and lifted off.  Special care is taken to deposit the Cr layer at 0.3A/s in order to improve coating uniformity.

\subsection{Measurement}
Experiments were performed in a dry dilution refrigerator with a base temperature of ~20mK.  Electronic filters are used in line with transport and gate contacts in order to lower the effective electron temperature.  To improve edge mode equilibration to the contacts most measurements are performed at 300mK unless otherwise noted.  Electronic measurements were performed using standard lock-in amplifier techniques.  For the diagonal conductance measurements an AC voltage bias at 17.77Hz is applied via a 1000x resistor divider to (see Fig.~\ref{fig:integer_qpc}a for contact references) C3-4 and the resulting current is measured using an Ithaco 1211 trans-impedance amplifier on C5-6 with a gain of $10^{-7}$ A/V.  We use transport contacts in pairs to decrease the contact resistance and improve edge-state equilibration, exciting an AC voltage on contacts C3/C4, and the measuring the current $I_{out}$ on contacts C5/C6. The diagonal voltage drop, $V_D$, is measured between C1/C2 and  C7/C8.  We then define $G_{\mathrm{D}} \equiv I_{out} / V_D$. Note that the sign convention used here is such that $G_D$ is positive for data measured on the hole side of the device, $\nu < 0$.
The voltage is measured between contacts C1-2 and C7-8 with an SR560 voltage pre-amplifier with a gain of 1000.  For two terminal measurements the same AC bias is applied to contacts C1-4 and the current is measured via C5-8.  DC bias was added on top of the AC bias using a passive summer.

\subsection{Thomas-Fermi calculation}

We consider a classical effective model wherein the electron density $n(\mathbf{r})$ of the two-dimensional electron gas adjusts according to the local electrostatic potential and compressibility of an interacting Landau-level (LL).
The classical energy functional can be decomposed into the Hartree energy, the interaction with an externally applied  potential $\Phi(\mathbf{r})$, and the remainder,
\be
    \begin{aligned}
        \mathcal{E}[n(\mathbf{r})] &= \frac{e^2}{2}\int_{{\bf{r_1}, \bf{r_2}}} n({\bf{r_1}}) V({\bf{r_1}, \bf{r_2}}) n({\bf{r_2}})\\
        & \qquad \qquad - e\int_{\mathbf{r}} \Phi(\mathbf{r}) n(\mathbf{r}) + \mathcal{E}_{xc}[n(\mathbf{r})],
    \end{aligned}
\ee
$\mathcal{E}_{xc}[n]$ contains not only the exchange-correlation energy but also single-particle contributions (e.g. inter-LL interactions, , Zeeman energies, etc.). If $n(\mathbf{r})$ varies slowly compared with $\ell_{B}$, we may neglect the dependence of the functional $\mathcal{E}_{xc}[n(\mathbf{r})]$ on the gradient $\boldsymbol{\nabla} n$ and employ the local density approximation (LDA),
\begin{equation}
    \mathcal{E}_{xc}[n(\mathbf{r})] = \int_{\mathbf{r}} E_{xc}(n(\mathbf{r}))
\end{equation}
where $E_{xc}(n)$ is determined for a system at \textit{constant} density $n$.
Our aim here is to find the density configuration $n(\mathbf{r})$ corresponding to the global minimum of the free energy $\mathcal{E}[n(\mathbf{r})]$.

We follow the geometry shown in Fig.~\ref{fig:fab_geometry}g. 
% The geometry we consider is shown in Fig.~\ref{fig:tf_model_summary}a-b.
There are four top gates a distance $d_t$ above the sample and one back gate a distance $d_b$ below, between which the space is filled by hBN with dielectric constant $\epsilon_{\perp} = 3$ and $\epsilon_{\|} = 6.6$ \cite{laturia_dielectric_2018}.
The N/S gates and the E/W gates are shifted by $w_{NS}$ and $w_{EW}$, respectively, from the center. 
% (see Fig.~\ref{fig:tf_model_summary}a-b).
We make an approximation by treating the cut-out ``X''-shaped region to define the gates as  a metal held at fixed voltage $V=0$ (rather than as a vacuum). 
This allows us to analytically solve for the electrostatic Green's function and gate-induced potentials without resorting to e.g. COMSOL simulations.

For a coarse-grained system with a finite resolution grid, the classical energy functional
becomes
\begin{equation}
    \begin{aligned}
	    \mathcal{E}[\{n(\mathbf{r})\}] &= \mathcal{E}_C + \mathcal{E}_{xc} + \mathcal{E}_{\Phi}\\
	    \mathcal{E}_C &= \frac{1}{2 A} \sum_{\mathbf{q}} V(\mathbf{q}) n(\mathbf{q}) n(-\mathbf{q}) \\
	    \mathcal{E}_{xc} &= \sum_{\mathbf{r}} E_{xc}(n(\mathbf{r})) \dif A,\\
	    \mathcal{E}_{\Phi} &= \sum_{\mathbf{r}} \Phi(\mathbf{r}) n(\mathbf{r}) \dif A
	\end{aligned}
	\label{eq:energy_functional}
\end{equation}
where $\dif A = \dif x \dif y$ is the grid area, $A$ is the total area, and $n(\mathbf{q}) = \sum_{\mathbf{r}} e^{-i \mathbf{q} \cdot \mathbf{r}} n(\mathbf{r}) \dif A$.
$\mathcal{E}_C$ is set by the gate-screened Coulomb interaction,
\begin{equation}
    V(\mathbf{q})= \frac{e^2}{4 \pi \epsilon_0 \epsilon_{\mathrm{hBN}}} \frac{4 \pi \sinh \left( \beta d_t|\mathbf{q}| \right) \sinh \left(\beta d_b|\mathbf{q}|\right)}{\sinh \left(\beta(d_t + d_b)|\mathbf{q}| \right) |\mathbf{q}|},
    \label{eq:gate_potential}
\end{equation}
$\epsilon_{\mathrm{hBN}} = \sqrt{\epsilon_{\perp} \epsilon_{\|}}$ and $\beta = \sqrt{\epsilon_{\|} / \epsilon_{\perp}}$. 
$\mathcal{E}_{\Phi}$ is the one body potential term arising from the potential $\Phi(\mathbf{r})$ on the sample due to the adjacent gates,
\begin{equation}
    \Phi(\mathbf{q}) = - e V_{t}(\mathbf{q}) \frac{\sinh(\beta d_b|\mathbf{q}|)}{\sinh(\beta(d_t + d_b)|\mathbf{q}|)} - e V_B \frac{d_t}{d_t + d_b} 
\end{equation}
where $V_{t}(\mathbf{q})$ is the top gate potential and $V_B$ is the back gate potential.

The remaining energy $E_{xc}$ is defined by integrating the chemical potential $\mu$,
\begin{equation}
    E_{xc}(n) = \int_0^{n} \dif n' \mu(n').
\end{equation}
which encodes information about the IQH and FQH gaps and the electron compressibility. For $\mu$ we use a phenomenological model detailed in SM.

Physically, $n(\mathbf{r})$ can only vary on the scale of the magnetic length $\ell_B$ due to the underlying quantum Hall wavefunction. 
To capture this feature, we implement a square grid with periodic boundary conditions and meshing much finer than the scale of $\ell_B$,
and then evaluate $\mathcal{E}$ with respect to the Gaussian convoluted density profile \footnote{We first use an unbounded range for $n(\mathbf{r})$ to determine the external potential that fix bulk filling $\nu_{EW} = 1$ and $\nu_{NS} = 0$. Then we constrain $n(\mathbf{r})$ to fall in between $\nu = 0$ and $\nu = 1$ such that any transition from $\tilde{\nu} = 0$ to $\tilde{\nu} = 1$ in $\tilde \nu(\mathbf{r})$ has a finite width of scale $\ell_B$.}
\begin{equation}
    \tilde n(\mathbf{r}) = \mathcal{N}^{-1}\sum_{\mathbf{r}'} n(\mathbf r') e^{-\abs{\mathbf{r} - \mathbf{r}'}^2 / 2\ell_B^2},
\end{equation}
where $\mathcal{N}$ is the corresponding normalization factor.
We use a basin-hopping global optimizer with local L-BFGS-B minimization to vary $\{n(\mathbf{r})\}$ and find the lowest energy configuration.

\section{Data Availability}
The data that support the findings of this study are available from the corresponding author upon reasonable request.

%\putbib[references]
%\end{bibunit}

\newpage
\clearpage

\onecolumngrid

\section{Extended Data Figures}
\begin{figure}[ht]
    \centering
    \includegraphics[width = 180mm]{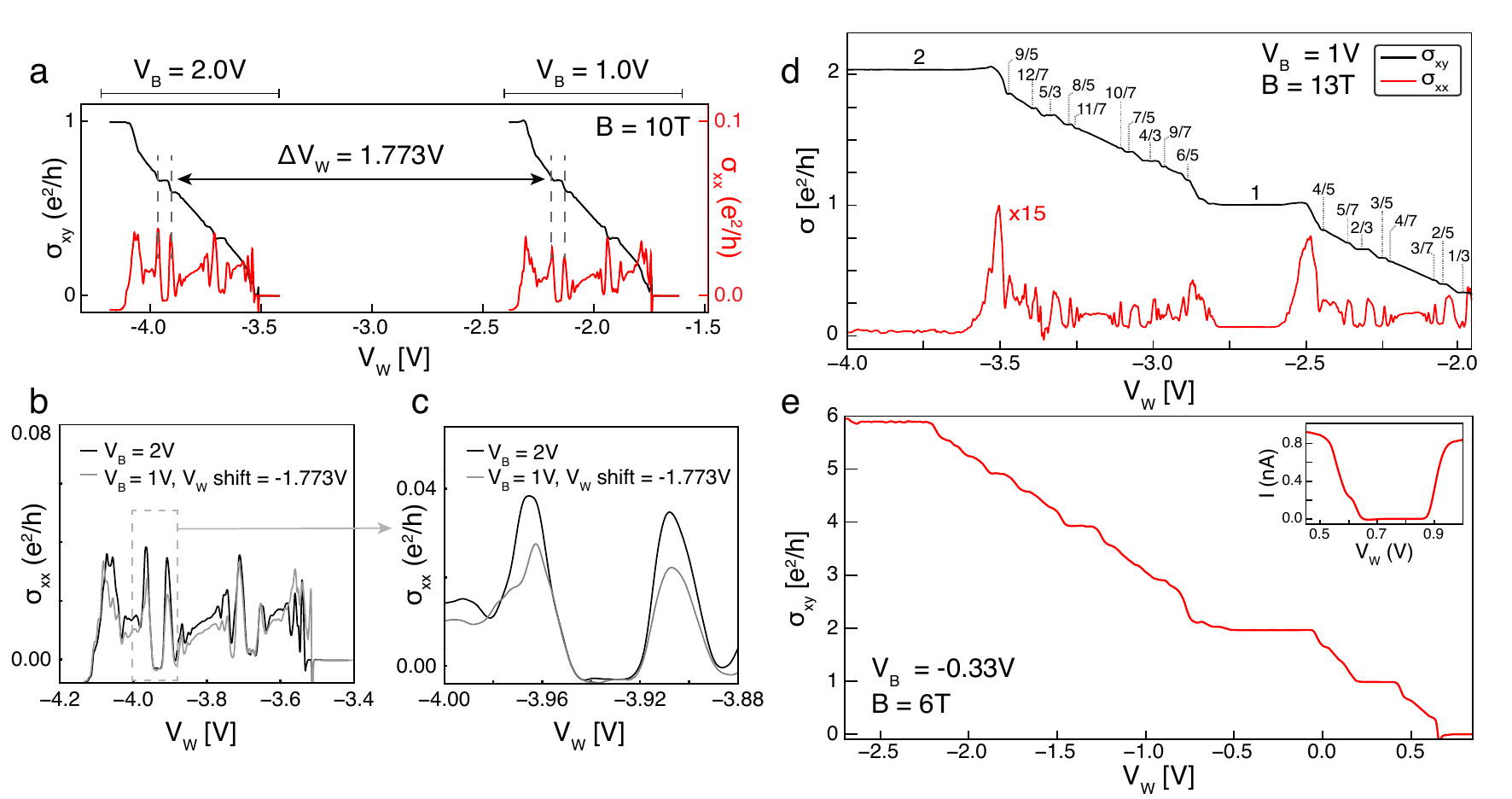}
    \caption{ \textbf{Precise Determination of $\alpha$ and Hall conductance at B = 6T and B = 13T:} \textbf{(a)} Longitudinal and Hall conductance in the W region of the device for two values of the bottom gate, $V_{\mathrm{B}}=1.0V$ and $V_{\mathrm{B}} = 2.0V$, at B = 10T.  All other regions are set to $\nu = 0$. \textbf{(b)} Calculated shift $\Delta V_W$ applied to the $V_{\mathrm{B}}=1.0V$ trace shows overlap of identical features between the two traces. \textbf{(c)} The region of vanishing longitudinal conductance in $\nu=-2/3$ was used to numerically determine the shift $\Delta V_W$ by minimizing the sum of the norm-squared differences between the two traces over a region around $\nu = -2/3$. \textbf{(d)} $\sigma_{xx}$ and $\sigma_{xy}$ versus $V_\text{W}$ at $B = 13T$ and $V_\text{B} = 1V$.  \textbf{(e)} $1/R_{xy}$ measured on the west side of the device versus $V_{W}$ while $\nu_{N} = \nu_{S} = \nu_{E} = 0$ is kept fixed, and $V_{\mathrm{B}} = -0.33V$.  Inset: current measured during $1/R_{xy}$ sweep showing $\nu = 0$ gap.}
    \label{fig:supp_determining_alpha}
\end{figure}

\begin{figure*}[htbp]
    \centering
    \includegraphics[width=0.9\textwidth]{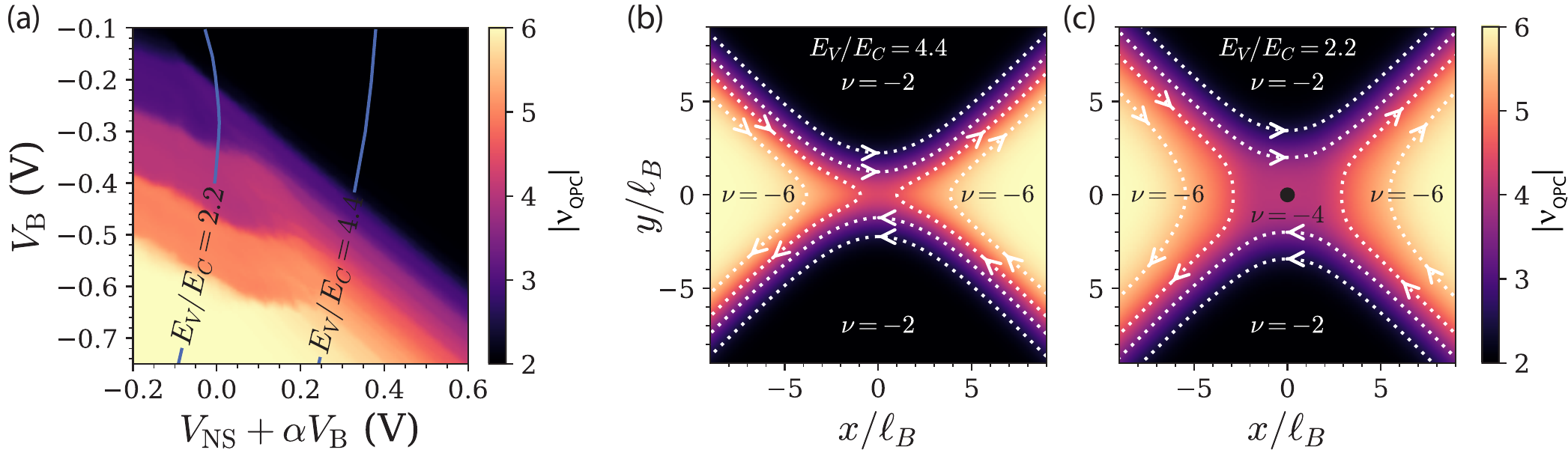}
    \caption{
    \textbf{Thomas-Fermi simulations of the QPC pinch-off}
    \textbf{(a)}
    The filling at the center of the QPC $\nu^{\mathrm{QPC}}$ as a function of $V_B$ and $V_{\text{NS}} + \alpha V_B$, with the bulk filling of the east/west regions fixed at $\nu_{EW} = -6$. The result qualitatively mimics the measured $G_D$ shown in Fig.~\ref{fig:2T_electrostatic}a since $\nu^{\mathrm{QPC}}$ determines the number of transmitted modes and therefore the diagonal conductance. Two line cuts which correspond to $E_V/E_C = 2.2$ and $4.4$ are shown in Fig.~\ref{fig:2T_electrostatic}c. \textbf{(b-c)} Calculated filling factor $\nu$ for a realistic device geometry at $B = \SI{2}{T}$ for (b) $E_V/E_C = 4.4$ and (c) $E_V/E_C = 2.2$. When $E_V/E_C = 2.2$, there exists an incompressible island with $\nu^{\textrm{QPC}} = -4$ at the center of the QPC. Contours of $\nu = n + 1/2$ are shown as white dotted lines, indicating the location of chiral edge modes, two-pairs of which are transmitted through the QPC. This illustrates the rule
    $N^{\textrm{QPC}} = \nu^{\textrm{QPC}} + 2$.}
    \label{fig:tf_model_summary_2}
\end{figure*}

\clearpage
\pagebreak

\renewcommand\thefigure{S\arabic{figure}}
\setcounter{figure}{0}

\title{Supplementary Information: Nanoscale electrostatic control in ultra clean van der Waals heterostructures by local anodic oxidation of graphite gates}
\maketitle
\onecolumngrid
\newpage 

\section{Partitioning of integer quantum Hall edge modes}

\begin{figure*}[ht]
    \centering
    \includegraphics[width = 180mm]{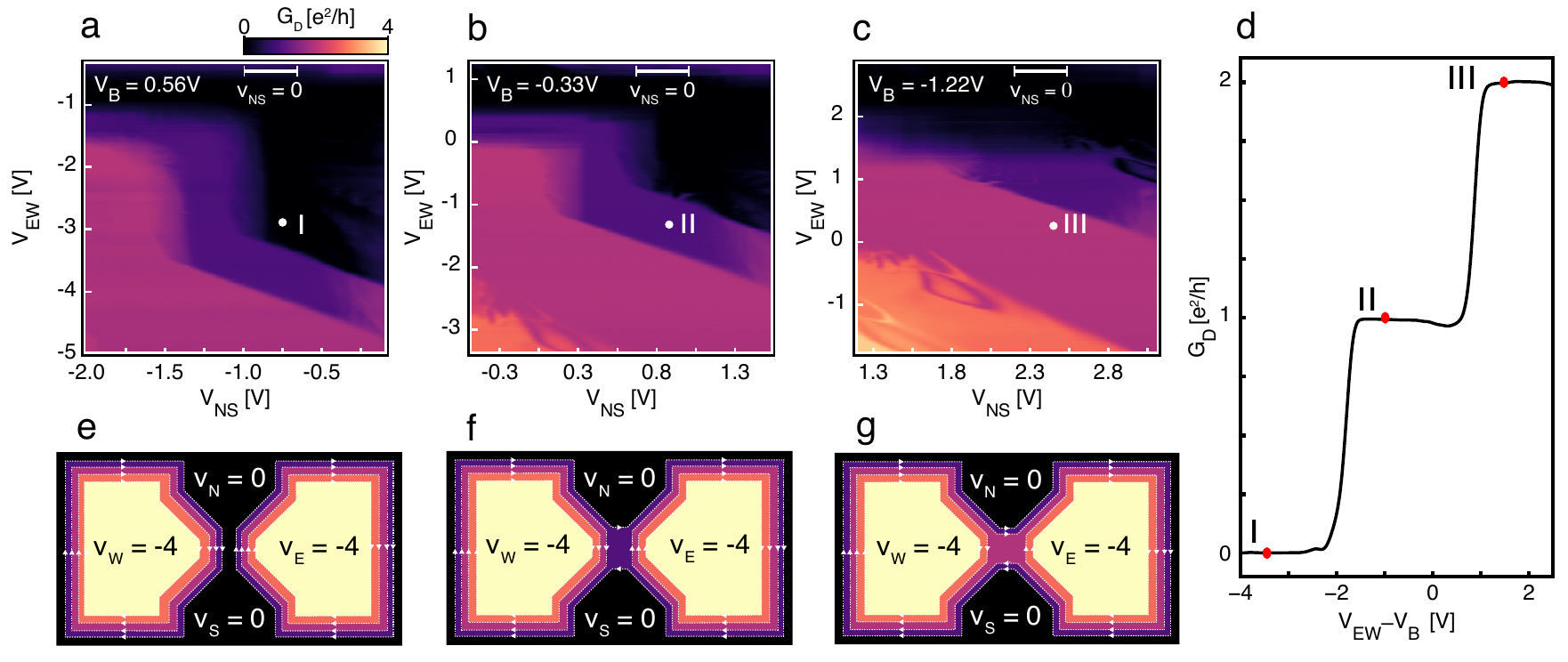}
    \caption{\textbf{Quantum point contact operation in the integer quantum Hall regime.} 
    \textbf{(a)} Diagonal conductance $G_{\mathrm{D}}$ plotted as a function of $V_{\text{NS}}$ and $V_{\text{EW}}$ at B=6~T and T=300~mK for $V_{\mathrm{B}}=0.56$,
     \textbf{(b)} $V_{\mathrm{B}}=-0.33$, and 
     \textbf{(c)} $V_{\mathrm{B}}=-1.22$. In panels a-c, $V_{\text{NS}}$ and $V_{\text{EW}}$ denote the voltages applied to north and south or east and west gates, respectively.  The ranges of $V_{\text{NS}}$ and $V_{\text{EW}}$ are chosen such that the filling factor $\nu_{EW} = \nu_{E} = \nu_{W} \in [-6, 0]$ and $\nu_{NS} = \nu_S = \nu_N \in [-2, 1]$ for each conductance map in a-c. The range over which $\nu_{NS}=0$ is marked in each plot by the white bar.  
     \textbf{(d)} The same trace as in the main text Fig.~\ref{fig:integer_qpc}b.  The trace in the multidimensional parameter space intersects the $G_{\mathrm{D}}$ maps of panels a-c at the position marked I, II, and III 
     \textbf{(e)} Schematic depiction of the filling factors within the QPC at point I in panel a, 
     \textbf{(f)} point II in panel b, and 
     \textbf{(g)} point III in panel c.   For points I-III, the filling factor in the N/S/E/W regions is constant but the fringe fields vary with $V_{\mathrm{B}}$, fully modulating transmission of the two outermost edge modes through the QPC.}
    \label{fig:supp_integer_qpc}
\end{figure*}

Fig.~\ref{fig:supp_integer_qpc}a shows a map of the $G_{\mathrm{D}}$ at B=6~T and T=300~mK as we vary the $V_{\text{NS}}\equiv V_N = V_S$ and $V_{\text{EW}} \equiv V_E = V_W$.  
As expected for IQH transport, the conductance maps in Fig.~~\ref{fig:supp_integer_qpc}a-c are dominated by regions of fixed conductance at integer multiples of $\frac{e^2}{h}$, corresponding to transmission of an integer number of chiral edge states across the device.  For each conductance map in Fig.~~\ref{fig:supp_integer_qpc}a-c, the graphite bottom gate, $V_{\mathrm{B}}$, is fixed to a different voltage.  Using the capacitive lever arm $\alpha\equiv C_B/C_T=1.773$ (see Fig.~\ref{fig:supp_determining_alpha}), the ranges of $V_{\text{EW}}$ and $V_{\text{NS}}$ are chosen such that the electronic density in each region of the monolayer is kept in the same range in all three panels. The precise mapping of gate voltages to $\nu$ is determined by measuring the Hall effect in the W quadrant (see Fig.~\ref{fig:supp_determining_alpha}).  

Tracing the behavior of the transitions between conductance plateaus reveals two distinct regimes.  In the first, plateau transitions are controlled by only $V_{\text{EW}}$ \textit{or} $V_{\text{NS}}$, producing steps in $G_{\mathrm{D}}$ along either horizontal or vertical lines in the $(V_{\text{NS}} , V_{\text{EW}})$ plane.  In this regime, the diagonal conductance is primarily determined by the number of edge states transmitted along the physical edge of the device, far from the QPC.  For example at point I in Fig.~~\ref{fig:supp_integer_qpc}a, the filling factor of the north and south regions, $\nu_{NS} = \nu_N = \nu_S$, is fixed to $\nu_{NS} = 0$, while the E/W regions are fixed to $\nu_{EW} = \nu_E = \nu_W = -4$.  Point I sits to the right of a \textit{vertical} transition, solely controlled by $V_{\text{NS}}$, where $G_{\mathrm{D}}$ goes from 0 to 1.  
Decreasing $V_{\text{NS}}$ starting at point I changes the filling $\nu_{NS}$ from 0 to -1, adding an edge mode in the N/S regions at the physical device boundary and increasing $G_{\mathrm{D}}$.  

In the second regime, the plateau transitions are influenced by \textit{both} $V_{\text{EW}}$ and $V_{\text{NS}}$, producing a step in $G_{\mathrm{D}}$ along lines of slope $\approx -1$ in the $(V_{\text{NS}},V_{\text{EW}})$ plane.  This behavior is expected when edge modes are transmitted through the center of the QPC, where they are equally sensitive to the fringe electric fields of each of the N/S/E/W gates.  Consider again point I which sits just above a \textit{diagonal} transition, where $G_{\mathrm{D}}$ goes from 0 to 1.  Near the transition, $\nu_{NS} = 0$ and no conduction across the device can occur along the etched boundary -- all current must be carried via edge modes through the QPC.  However, at the transition, $G_{\mathrm{D}}$ may change sharply by \textit{either} an equal perturbation in $V_{\text{NS}}$ or $V_{\text{EW}}$ while maintaining $\nu_{NS} = 0$.  The existence of such transitions in $G_{\mathrm{D}}$ implies the filling factor in the QPC center can be held fixed via equal and opposite modulations of $V_{\text{NS}}$ and $V_{\text{EW}}$: lines separating differing values of $G_{\mathrm{D}}$ which are parallel to $V_{\text{NS}} + V_{\text{EW}} = 0$ demarcate sharp boundaries between regions of different filling factor in the center of the QPC itself.   

The location in density of the \textit{diagonal} steps in $G_{\mathrm{D}}$  shift as a function of $V_{\mathrm{B}}$ (Figs.  \ref{fig:supp_integer_qpc}b-c), in contrast to the horizontal and vertical transitions whose locations in density are unaffected. This behavior follows from the device electrostatics: Near the device boundary, transport is determined directly by the bulk filling factor in the N/S/E/W regions.  Since the graphite bottom gate uniformly modulates the density of the whole monolayer, the role of $V_{\mathrm{B}}$ is merely to induce a chemical potential shift in the whole device which is compensated by offsetting the applied gate voltages.  This is not true in the central region, however, which is doped by the fringe fields of N/S/E/W gates.

It follows that tuning the bottom gate while keeping the densities in the N, S, E, and W regions constant changes the electrostatics of the QPC. 
Points I-III in Figs. \ref{fig:supp_integer_qpc}a-c correspond to identical carrier densities away from the QPC, with $\nu_{EW} = -4$ and $\nu_{NS} = 0$. 
At point I, all the edge modes are pinched off and $G_{\mathrm{D}} = 0$. As $V_{\mathrm{B}}$ is decreased, the filling factor in the QPC changes, leading to the transmission of one additional edge modes at point II and and two additional modes at point III.  
Fig.~\ref{fig:supp_integer_qpc}d shows the continuous evolution between points I, II, and III as a function of $V_{\text{EW}} - V_{\mathrm{B}}$, corresponding to tuning the QPC electrostatics. A schematic depiction of the corresponding filling factor maps in real space are shown in Figs. \ref{fig:supp_integer_qpc}e-g.

\section{Extended Integer QPC Operation}

\onecolumngrid

\begin{wraptable}{r}{5.5cm}
\begin{tabular}{||c c c c c||}
 \hline
 \textbf{Index} & $G_{\mathrm{D}} [\frac{e^2}{h}]$ & \textbf{$\nu_{EW}$} & \textbf{$\nu_{NS}$} & \textbf{$\nu_{qpc}$} \ \\ [0.8ex] 
 \hline\hline
 I & 1 & -4 & 0 & -1 \\ 
 \hline
 II & 1 & -2 & -1 & -1 \\
 \hline
 III & 1 & -1 & -2 & 0 \\
 \hline
 IV & 1 & -1 & -2 & -1 \\
 \hline
 V & 2 & -2 & -2 & -2 \\ 
 \hline
 VI & 0.7 & -1 & -4 & -2 \\ [1ex] 
 \hline
\end{tabular}
\end{wraptable}

\begin{figure*}[hb]
    \centering
    \includegraphics[width = 170mm]{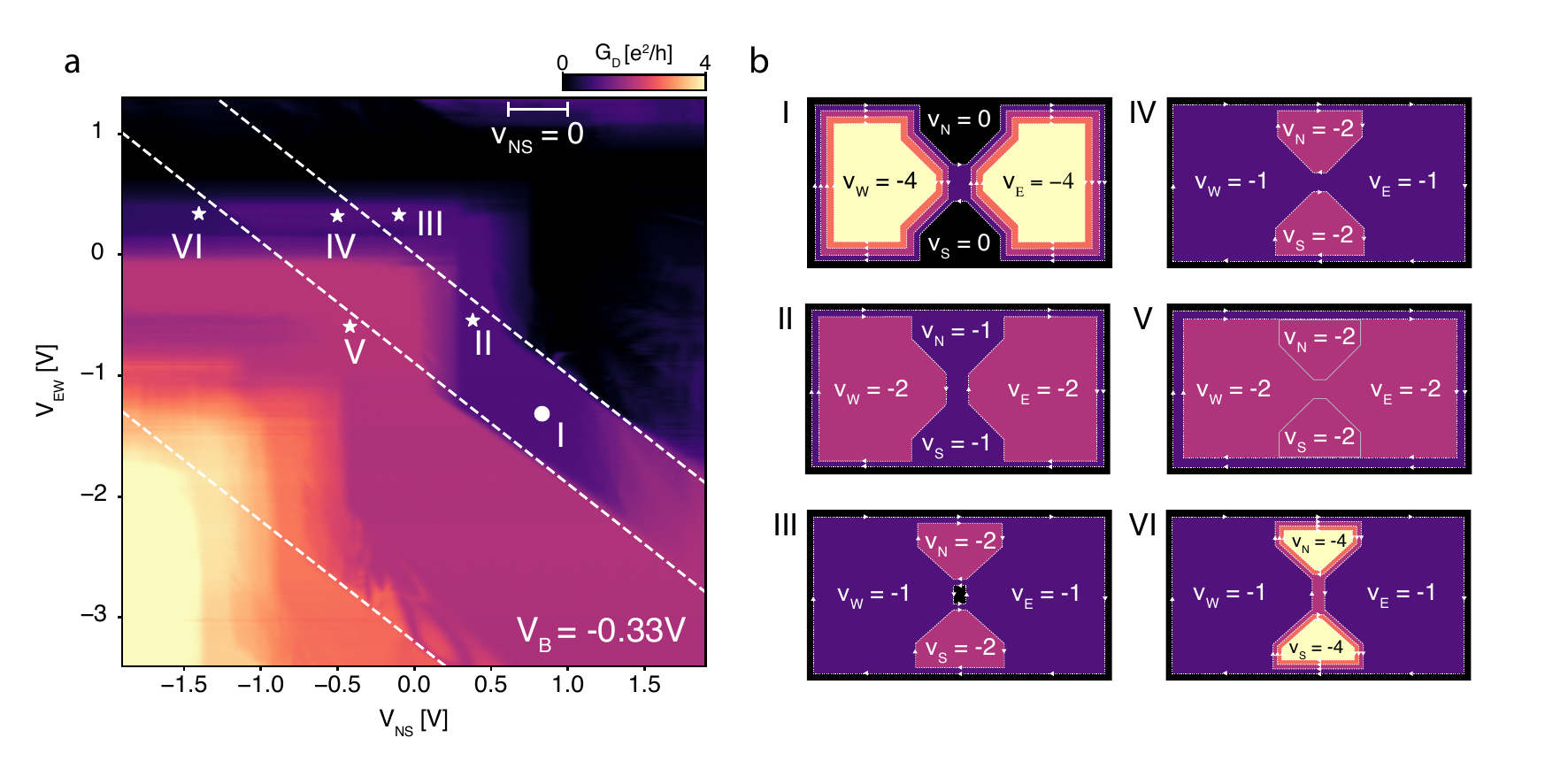}
    \caption{\textbf{Extended Integer QPC Operation:} \textbf{(a)} $G_{\mathrm{D}}$ plotted versus $V_{\text{NS}}$ and $V_{\text{EW}}$ with $V_{\mathrm{B}} = -0.33V$.  Lines parallel to $V_{\text{NS}} + V_{\text{EW}} = 0$ demarcate transitions between integer $G_{\mathrm{D}}$ values indicating a rapid change in the density at the QPC.  Roman numerals I-VI correspond to various unique combinations of $\nu_{NS}$, $\nu_{EW}$, and $\nu_{qpc}$, where $\nu_{qpc}$ is the filling factor at center of the QPC.  \textbf{(b)} Illustrations of the inferred filling factor in each region of the device corresponding to the operating points in panel a.}
    \label{fig:supp_ext_integer_qpc}
\end{figure*}

The full parameter space that determines the value of $G_{\mathrm{D}}$ when tuned with $V_{\text{NS}}$ and $V_{\text{EW}}$ has features which depend on either $V_{\text{NS}}$ or $V_{\text{EW}}$ individually or features perpendicular to $V_{\text{NS}} + V_{\text{EW}} = 0$ (indicated as dashed white lines in Fig.~\ref{fig:supp_ext_integer_qpc}a).  The former are dominated by physics at the etched edge of the device where the edge modes of a $p p' p$ junction are fully equilibrated \cite{zimmermann_tunable_2017}.  The latter are interpreted as boundaries between operating points that have differing values of $\nu_{qpc}$ -- the filling factor in the center of the device, determined by the fringe fields of the four quadrant top gates (for a fixed $V_{\mathrm{B}}$).  Fig.~\ref{fig:supp_ext_integer_qpc}a has several points denoted by roman numerals which correspond either to QPC operation or conductance through the edge of the device.  Fig.~\ref{fig:supp_ext_integer_qpc}b shows an illustration of the inferred filling factor in each region of the device for each point marked in Fig.~\ref{fig:supp_ext_integer_qpc}a.  This information, along with the associated $G_{\mathrm{D}}$ for each point is repeated in the table. 

Point VI is of particular note since it falls to the left of a boundary in $G_{\mathrm{D}}$, where $G_{\mathrm{D}}$ is a fractional value below 1, that is intersected by a dashed line.  This indicates the transition is sensitive to the potential at the QPC, but is in a region where at least one edge mode is transmitted along the edge of the device since $\nu_{NS} < -1$.  The corresponding filling factors at VI are $\nu_{NS} = -4$, $\nu_{EW} = -1$, and $\nu_{qpc} = -2$.  Starting from point IV, where $\nu_{qpc} = -1$, as $V_{\text{NS}}$ becomes more negative, the filling factor in the QPC is more strongly doped towards $\nu_{NS}$.  Eventually, at point VI, the filling factor in the QPC increases to $\nu_{qpc} = -2$, and an edge mode bridges the north and south regions through the $\nu_{EW} = -1$ bulk.  This creates a scattering channel across the device that reduces the conductance to the expected value of $G_{\mathrm{D}} \sim 2/3$ for a $p p' p$ junction of the given filling factors (-1, -2, -1) \cite{zimmermann_tunable_2017, deprez_tunable_2021}.  This shows it is possible to see QPC behavior even with a background conductance through the etched edge of the device, however the value of $G_{\mathrm{D}}$ requires some interpretation.  Consequently, for most experiments presented in the main text we focus exclusively on regimes where there is no conduction along the edge of the device, simplifying possible interpretations of fractional values of $G_{\mathrm{D}}$.

\section{DC Voltage Bias Correction}

We used a set of cryogenic filters at the mixing chamber in order to help equilibrate the electrons in the graphene 2DEG to the temperature of the dilution refrigerator (at the mixing chamber).  These filters contain a set of RF pi-filters with small component values which at low meausrement frequencies are negligible.  In addition, there are a set of RC filters with a roll-off frequency of 6.6kHz; these filters have a series resistance of $R = \SI{3}{k\Omega}$ each. In order to investigate the DC bias directly across the QPC it is necessary to take into account known filter resistances since they inherently form a voltage divider between themselves and the two-terminal resistance across the QPC defined as $R_{\text{DUT}}$. Determining the relationship between the applied DC voltage at the input of the filters (a known value defined as $V_{\text{DC}}$) and the voltage bias across the sample, defined as $V_{\text{DUT}}$, is straightforward.  At low frequencies, the current drawn to any floating contacts is negligible so the DC voltage across the sample is well approximated by 
\begin{equation}
    V_{\text{DUT}} = V_{\text{DC}} - (2(R/2) + 50\Omega) \cdot I
\end{equation}
where $I$ is the DC current through $R_{\text{DUT}}$.  In all tunneling measurements performed in this work pairs of contacts are tied together, so the effective series filter resistance is $R/2$.  The $\SI{50}{\Omega}$ is the lock-in output impedance. While the DC current is not known a priori, differentiating with respect to $V_{\text{DUT}}$ the above equation gives:
\begin{equation}
    1 = \frac{\mathrm{d} V_{\text{DC}}}{\mathrm{d} V_{\text{DUT}}} - (R + 50\Omega) \frac{\mathrm{d} I}{\mathrm{d} V_{\text{DUT}}}
\end{equation}
This equation can be rearranged to isolate $\mathrm{d} V_{\text{DUT}}$ and $\mathrm{d} V_{\text{DC}}$, given below.
\begin{equation}
    \mathrm{d} V_{\text{DUT}} = \frac{\mathrm{d} V_{\text{DC}}}{1 + (R + 50\Omega) \frac{\mathrm{d} I}{\mathrm{d} V_{\text{DUT}}}} 
\end{equation}
The differential conductance across the QPC, $G_{\mathrm{D}} = \mathrm{d} I/ \mathrm{d} V_{\text{DUT}}$, appears in the denominator.  Integrating this equation leaves the final correction formula, which can be expanded for $3050 \Omega \cdot G_{\mathrm{D}} \ll 1$ to give:
\begin{equation}
    V_{\text{DUT}} =  \int_0^{V_{\text{DC}}} \frac{\mathrm{d} V_{\text{DC}}'}{1 + (R + 50\Omega) G_{\mathrm{D}} (V_{\text{DC}}')} \approx V_{\text{DC}} - (3050\Omega) \int_0^{V_{\text{DC}}} \mathrm{d} V_{\text{DC}}' G_{\mathrm{D}}(V_{\text{DC}}')
\end{equation}
This correction is applied for all the tunneling curves in the main text Fig.~\ref{fig:fractional_qpc}d and the collapsed curves in Fig.~\ref{fig:fractional_qpc}f.

\section{Luttinger Tunneling Data Analysis}

\begin{figure*}[ht]
    \centering
    \includegraphics[width = 178mm]{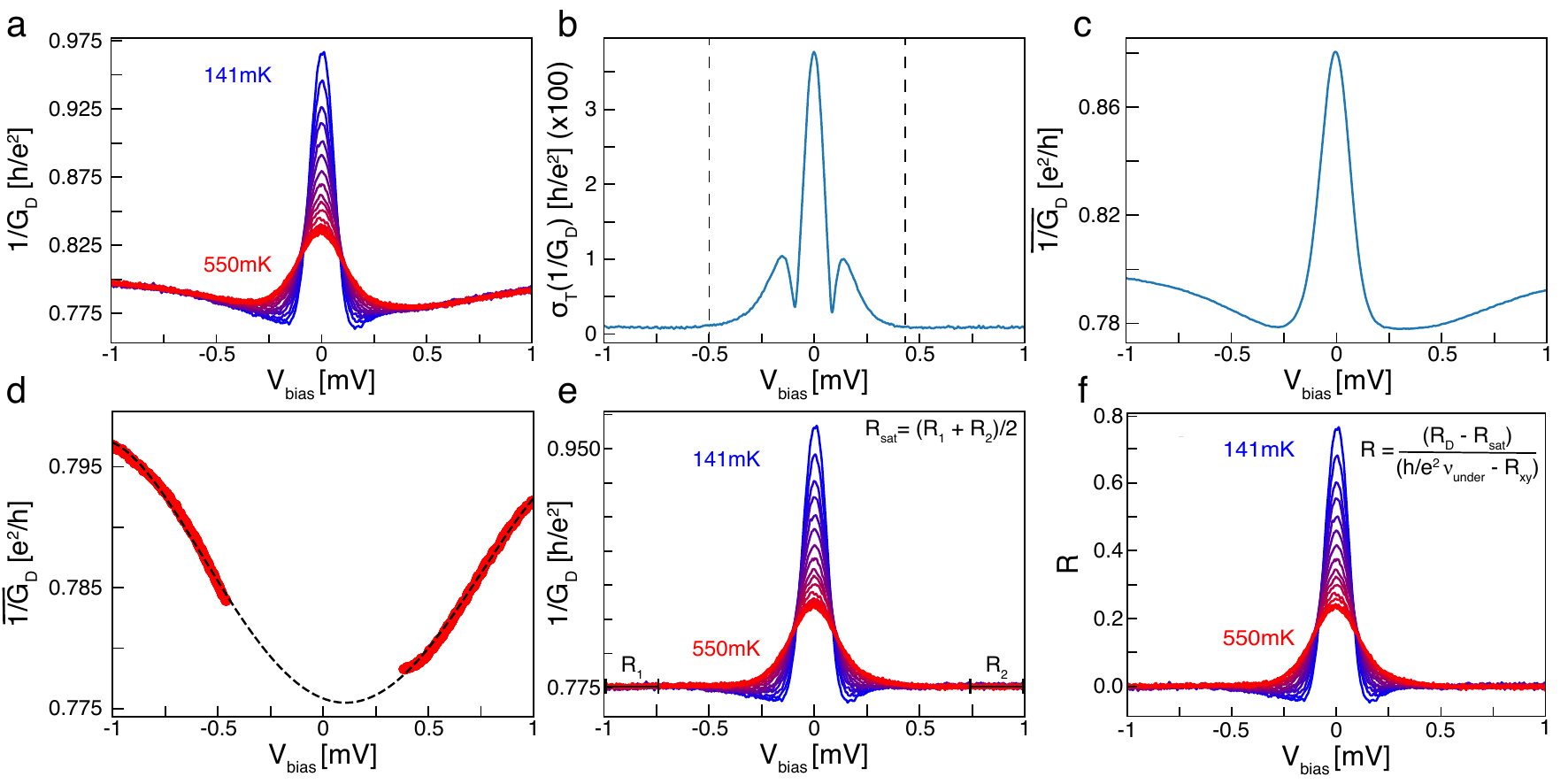}
    \caption{\textbf{Extended analysis of temperature dependent tunneling data} \textbf{(a)} Diagonal resistance, defined as $1/G_{\mathrm{D}}$, versus $V_{\text{bias}}$ for the same gate configuration as Fig.~\ref{fig:fractional_qpc}c at $V_{\text{NS}} = -3.925V$ and same temperature values as Fig.~\ref{fig:fractional_qpc}d.  \textbf{(b)} The standard deviation of $R_D$ defined as $\sigma R_D$ versus bias taken across all temperatures between $T = \SI{141}{mK}$ and $T = \SI{550}{mK}$.  Additional temperature values at finer increments are included which are not shown in panel (a).  Dashed lines indicate threshold in $V_{\text{bias}}$ where $\sigma R_D < 0.0012$.  \textbf{(c)} $R_D$ versus $V_{\text{bias}}$ averaged over all temperatures.  \textbf{(d)} Portion of the temperature averaged $R_D$ in panel (c) plotted only where the standard deviation $\sigma R_D$ falls below $0.0012$.  The black dashed line is a fit to a quartic polynomial which then defines the temperature independent background.  \textbf{(e)} $R_D$ plotted versus $V_{\text{bias}}$ for various temperatures with the quartic background extracted in panel d subtracted (except for the constant term).  A value for $R_{\text{sat}}$, i.e, the resistance which panel (e) saturates to can be defined by averaging $R_D$ over a window of $\SI{250}{\mu V}$ on either side of $V_{\text{bias}} = 0$ at high bias, and then averaging together the results. \textbf{(f)} $R$ plotted versus $V_{\text{bias}}$ for the same temperatures in Fig.~\ref{fig:fractional_qpc}d. }
    \label{fig:supp_tunneling_analysis}
\end{figure*}

The theoretical expectation given by Eq.~\eqref{eqn:digamma} is that the reflection coefficient $R$ has no terms which depend solely on voltage bias and not temperature.  Consequently, to compare our raw data with the predictions from chrial Luttinger liquid theory we need to remove a temperature independent background.  Fig.~\ref{fig:supp_tunneling_analysis}a shows the unprocessed dependence of $1/G_{\mathrm{D}}$ versus $V_{\text{bias}}$ for the same series of temperatures listed in Fig.~\ref{fig:fractional_qpc}d.  At large voltage bias, the temperature dependent part of $1/G_D$ is expected to flatten out, however while we do observe negligible temperature dependence for $|V_{\text{bias}}| > \SI{500}{\mu V}$, there is a smooth variation in the bias dependence which is the same for all temperatures.  Fig.~\ref{fig:supp_tunneling_analysis}b shows the standard deviation of $1/G_D$ over temperature, denoted as $\sigma_T (1/G_D)$, versus $V_{\text{bias}}$.  We define the voltage bias value for which there is no longer any discernible temperature variation between the data sets by choosing $V_{\text{bias}}$ such that $\sigma_T (1/G_D) < 0.0012$.  This range includes all values of $V_{\text{bias}}$ which fall outside the bounds of the black dashed lines in Fig.~\ref{fig:supp_tunneling_analysis}b.  Fig.~\ref{fig:supp_tunneling_analysis}c shows $1/G_D$ versus $V_{\text{bias}}$ averaged over all data sets at varying temperatures between $\SI{141}{mK}$ and $\SI{550}{mK}$, defined as $\overline{1/G_D}$ (additional data sets not shown in Fig.~\ref{fig:supp_tunneling_analysis}a for clarity are included in this average).  The background, plotted in Fig.~\ref{fig:supp_tunneling_analysis}d, is then defined by $\overline{1/G_D}$ where ever $\sigma_T (1/G_D) < 0.0012$; the black dashed line in the same panel is the best-fit fourth order polynomial to the extracted background.  Fig.~\ref{fig:supp_tunneling_analysis}e shows $1/G_D$, for the same temperatures in Fig.~\ref{fig:supp_tunneling_analysis}a, with the background polynomial subtracted (not including the constant offset).  Here we define the saturation resistance $R_{\text{sat}}$ as the value $\overline{1/G_D}$ saturates to at high bias for all temperatures; practically this is computed by averaging $\overline{1/G_D}$ over $V_{\text{bias}} \in [-\SI{1}{mV}, \SI{-0.75}{mV}]$ and $V_{\text{bias}} \in [\SI{0.75}{mV}, \SI{1}{mV}]$, yielding an $R_{\text{sat}} = 0.775 \; h/e^2$.  Knowing $R_{\text{sat}}$, the value of the reflection coefficient can be computed from $R_D \equiv 1/G_D$, given by the formula inset into Fig.~\ref{fig:supp_tunneling_analysis}f, following the procedure outlined in Refs. \onlinecite{wen_edge_1991,roddaro_nonlinear_2003,radu_quasi-particle_2008,baer_experimental_2014}.

\section{Thomas-Fermi calculation of the local density across the edge}

\begin{figure*}[htbp]
    \centering
    \includegraphics[width=\textwidth]{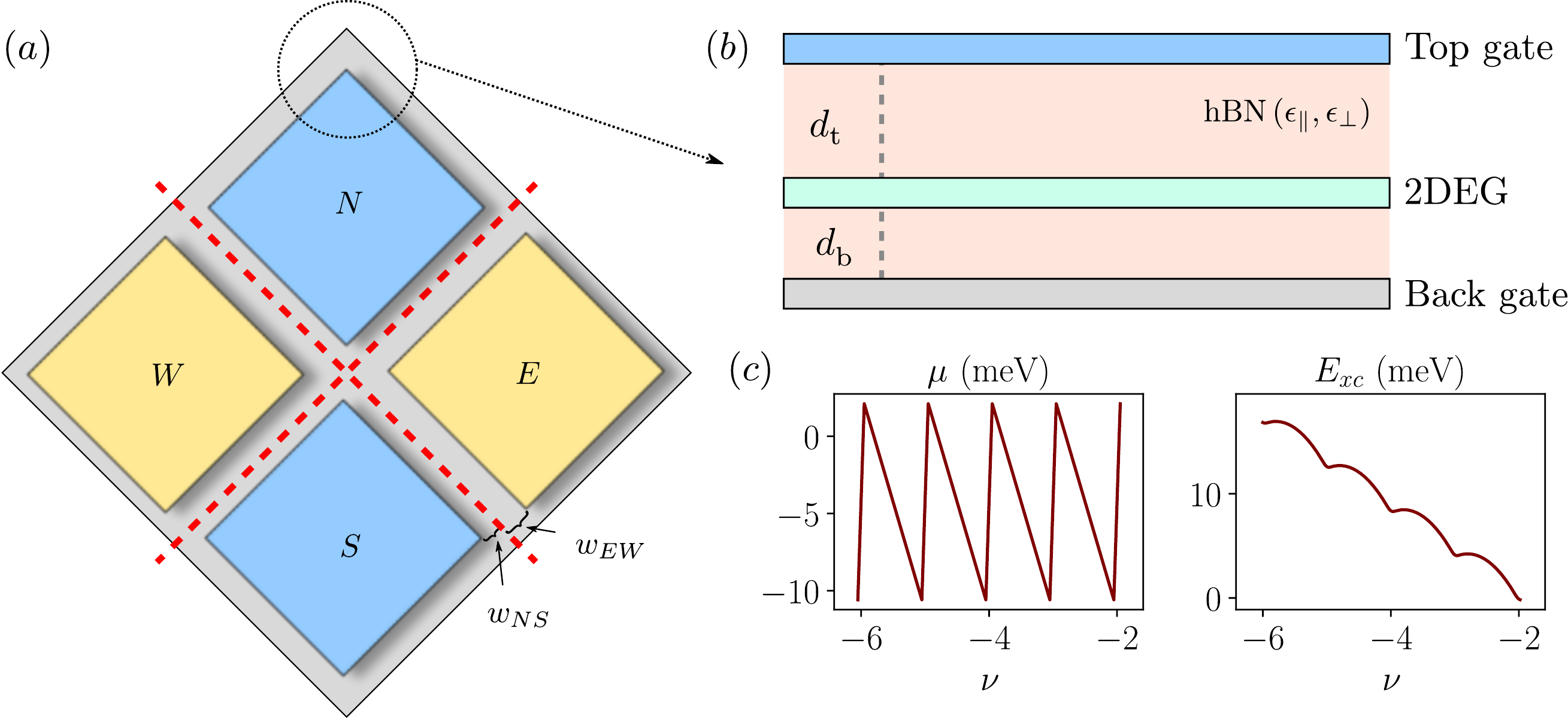}
    \caption{
    \textbf{Effective model considered within the Thomas-Fermi framework.}
    \textbf{(a)} Top-view of the gate setup, which involves four square gates labeled N,S,E, and W respectively, each of which is held at a potential denoted by $V_{N,S,E,W}$.
    The back-gate (gray) is held at constant $V_B$.
    We allow for the gate displacement from the center to be different for E/W and N/S gates.
    \textbf{(b)}
    Side-view of the gates reveals two distances $d_t$ and $d_b$ for the separation between the 2DEG and the top gate or bottom gate, respectively.
    \textbf{(c)} A phenomenological model of the chemical potential and the corresponding internal energy $E_\textrm{xc}$ considered in SM~\ref{sec:phenon_mu}.
    }
    \label{fig:tf_model_summary}
\end{figure*}

\subsection{Phenomenological model of the chemical potential} \label{sec:phenon_mu}
Here we provide details of the Thomas-Fermi calculations for the QPC with $\nu_{EW} = -6, \nu_{NS} = -2$, $B = \SI{2}{T}$.
Thomas-Fermi calculations require as input the filling-dependent chemical potential $\mu(\nu)$, as has been measured recently \cite{yang_experimental_2021}.
However, those measurements were at high-field as opposed to $B=\SI{2}{T}$. 
We thus construct a phenomenological model for $\mu(\nu)$ which captures the dominant low-field features: jumps in the chemical potential at the IQH gaps, and negative compressibility in between.
In monolayer graphene (MLG), the single-particle integer quantum Hall (IQH) gaps appear at $\nu = \pm 2, \pm 6, \cdots$. Within the parameter regimes in Fig.~\ref{fig:2T_electrostatic}a, we focus on the Landau-level (LL) $N = -1$ between $\nu = -6$ and $-2$ and assume the IQH gap  below $\nu = -6$ and above $\nu = -2$ is effectively infinite.
As we will apply our model at low field ($B = \SI{2}{T}$) we  neglect the fine structure coming from fractional quantum Hall (FQH) gaps in between. 
The dominant features in $\mu(\nu)$ are then (1) jumps in the chemical potential $\Delta \mu$ at integer fillings $\nu = -5$, $-4$ and $-3$ due to quantum Hall ferromagnetism (QHF) in the isospin space $(K, \uparrow), (K, \downarrow), (K', \uparrow), (K', \downarrow)$; (2) a compensating negative-compressibility $\frac{\partial \mu}{\partial\nu} \approx - \Delta \mu$ at intermediate fillings. 

The thermodynamic gaps $\Delta \mu$ of the QHFs contain a single particle contribution from the valley and spin splittings $E_V, E_Z$, and an interacting contribution from Coulomb exchange. 
In an $N\neq 0$ LL, we expect $E_V \approx 0$ (this is in contrast to the $N=0$ LL, where valley is locked to sublattice, and a sublattice splitting is generated by the hBN substrate). 
 The Zeeman energy $E_Z = \SI{0.23}{\meV}$ is also very small at $B = \SI{2}{T}$. 
 The quantum Hall ferromagnet gap $\Delta \mu$ is thus dominated by the Coulomb exchange contribution \cite{yang_experimental_2021} 
\begin{equation}
    \Delta \mu = \int \frac{d \mathbf{q}^2}{(2 \pi)^2} V(\mathbf{q}) |F(\mathbf{q})|^2 = \SI{12.7}{\meV}
\end{equation}
where $V(\mathbf{q})$ is the gated screened Coulomb interaction as in Eq.~\eqref{eq:gate_potential} and $F(\mathbf{q}) = ( F_{00}(\mathbf{q}) + F_{11}(\mathbf{q}))/2$ is the MLG form factor in the $N = -1$ LL. 
\begin{equation}
    F_{m, n}(\mathbf{q}) =e^{-\frac{1}{4} |\mathbf{q}|^2 \ell_B^2}\left(\frac{i |\mathbf{q}| \ell_B}{\sqrt{2}}\right)^{m-n} L_n^{m-n}\left(\frac{|\mathbf{q}|^2}{2}\right) \sqrt{\frac{n !}{m !}}
\end{equation}
is the usual GaAs form factor with $L_b^a(|\mathbf{q}|)$ being the generalized Laguerre polynomial.

Based on $E_V = E_Z \approx \SI{0}{\meV}$ and $\Delta \mu = \SI{12.7}{\meV}$, we consider a phenomenological model of the chemical potential as shown in Fig.~\ref{fig:tf_model_summary}c. We phenomenologically include the effect of disorder broadening such that $\mu$ increases by $\Delta \mu$ within a small window $\delta \nu = \pm 0.005$ close to  integer filling. 

\subsection{Evolution of the QPC pinch-off}

To approximate the experimental device, we simulate a system with magnetic field $B = \SI{2}{T}$ (magnetic length $\ell_B = \SI{18.14}{nm}$), gate distances $d_t = \SI{60}{nm}, d_b = \SI{30}{nm}$ and channel widths $w_{EW} = w_{NS} = \SI{31}{nm}$ (see Fig.~\ref{fig:tf_model_summary_2}a for definition) on a $30 \ell_B \, \times \, 30 \ell_B$ system with grid size $\dif x = \dif y = \ell_B/2$. We numerically checked that the charge density has converged in both system size and grid size.

As explained in Fig.~\ref{fig:tf_model_summary}a, the quantity we are most interested in from the Thomas-Fermi calculation is the charge density $\nu^{\mathrm{QPC}}$ at the center of the QPC, which determines the number of edge modes $N^{\mathrm{QPC}}$ transmitted through the QPC. $N^{\mathrm{QPC}}$ can be directly related to the total number of edge modes $N$ across the device and therefore the diagonal conductance $G_D$,
\begin{equation}
    N^{\text{QPC}} =  \nu^{\mathrm{QPC}} - (-2), \quad N = -2 + N^{\text{QPC}}, \quad G_{\mathrm{D}} \sim |N| \frac{e^2}{h} \sim \left |\nu^{\mathrm{QPC}} \right| \frac{e^2}{h},
\end{equation}
since there are two outer chiral modes outside the N/S region. We refer readers to Fig.~\ref{fig:supp_integer_qpc} and Fig.~\ref{fig:supp_ext_integer_qpc} for a more comprehensive illustration. 
Because the edge modes straddle the density contours $\nu \sim n + 1/2$, when $\nu^{\mathrm{QPC}}$ is not an integer certain edge modes approach very close to the saddle point,  leading to partial transmission and therefore a non-quantized diagonal conductance. 
Thus away from integer $N^{\text{QPC}}$ the identification 
$G_{\mathrm{D}} \sim N^{\text{QPC}} \frac{e^2}{h}$ is purely heuristic, as the Thomas-Fermi calculations cannot be expected to capture the transmission coefficient through the QPC.

In Fig.~\ref{fig:tf_model_summary_2}b, we show $\nu^{\mathrm{QPC}}$ as a function of gate voltages. As we pinch off the QPC, there is a clear transition from $\nu^{\mathrm{QPC}}$ dropping in a single sharp step to taking multiple steps due to the competition between the confinement energy $E_V$ and the Coulomb energy $E_C$. This transition indicates two  different charge density configurations within the QPC. When the confining potential is soft, i.e. $E_V/E_C$ is small, it is energetically more favorable to form an incompressible island at the QPC, so that $\nu^{\mathrm{QPC}}$ will quantize at integer fillings. 
When the confining potential is sharp, i.e. $E_V/E_C$ is large, it is energetically more favorable for the charge density $\nu^{\mathrm{QPC}}$ at QPC to change continuously. We note that the $E_V/E_C$ obtained the simulation is slightly larger than obtained in the experiment, which could originate from the screening from virtual inter-LL transitions that had been neglected in the simulation.

We show two prototypical charge density configurations in Fig.~\ref{fig:tf_model_summary_2}b-c corresponding to these two cases. When $E_V/E_C$ is large, charge density within the the QPC changes continuously as shown in Fig.~\ref{fig:tf_model_summary_2}b. When $E_V/E_C$ is small, a small incompressible island with $\nu  = -4$ forms at the QPC as shown in Fig.~\ref{fig:tf_model_summary_2}c. Whether this island extends to incompressible strips along the electrostatically defined edge depends on $E_V/E_C$ across the edge, which is generally expected to be larger since the confining potential is sharper at the edge than at the junction. In this IQHE setting, the edge structure far from the QPC is not so relevant to the diagonal conductance $G_D$ measured in the experiment. 

%TC:endignore
\end{document}

%"Edge Reconstruction and Localization of Fractional Charge in Graphene Quantum Hall Point Contacts"?
\title{Supplemental Information}
\author{Liam Cohen}   
\email{liamcohen@ucsb.edu}
\affiliation{Department of Physics, University of California at Santa Barbara, Santa Barbara CA 93106, USA}
\author{Noah Samuelson}   
\affiliation{Department of Physics, University of California at Santa Barbara, Santa Barbara CA 93106, USA}
\author{Takashi Taniguchi}
\affiliation{International Center for Materials Nanoarchitectonics,
National Institute for Materials Science,  1-1 Namiki, Tsukuba 305-0044, Japan}
\author{Kenji Watanabe}
\affiliation{Research Center for Functional Materials,
National Institute for Materials Science, 1-1 Namiki, Tsukuba 305-0044, Japan}
\author{Andrea F. Young}
\email{andrea@physics.ucsb.edu}
\affiliation{Department of Physics, University of California at Santa Barbara, Santa Barbara CA 93106, USA}

\date{\today}

\maketitle

\begin{figure}[ht]
    \includegraphics[width = 90mm]{supp_zerofield_splitgates.pdf}
    \caption{ Effect of the N/S gates at B=0T. The two terminal resistance measured across the device from W to E is shown as the global graphite bottom gate and N/S gates are swept. The E/W gates are fixed at $V_{EW}=0$. The horizontal line just above $V_{bg} = 0$ shows an order-of-magnitude increase in resistance when the global graphite bottom gate sets the density in the E/W regions near charge neutrality. When the N/S gates are tuned to charge neutrality, corresponding to the diagonal line, a sharp but less resistive peak is seen when a narrow constriction of conductance is formed between the two highly resistive regions under the N and S gates }
    \label{fig:supp_zerofield_splitgates}
\end{figure}

\begin{figure}[ht]
    \includegraphics[width = 90mm]{supp_repeated_resonances.pdf}
    \caption{ Repetition of identical peak and dip structure near each step in conductance across the QPC}
    \label{fig:supp_repeated_resonances}
\end{figure}

\section{Precise Determination of the Capacitance Ratio}

\begin{figure*}
    \centering
    \includegraphics[width = 180mm]{supp_determining_alpha.pdf}
    \caption{ Precise Determination of $\alpha$. \textbf{(a)} Longitudinal and Hall conductance of the W region of the device for two values of the back gate, $V_{bg}=1.0V$ and $V_{bg} = 2.0V$. \textbf{(b)} Calculated shift $\Delta V_W$ applied to the $V_{bg}=1.0V$ trace shows overlap of identical features between the two traces. \textbf{(c)} The region of vanishing longitudinal conductance at in $\nu=-2/3$ was used to programatically determine the shift $\Delta V_W$ by minimizing the norm-squared difference between the two traces over a region around $\nu = -2/3$. }
    \label{fig:integer_qpc}
\end{figure*}

Knowledge of the precise ratio of the capacitance per unit area to graphite top and bottom gates, namely $\alpha \equiv \frac{c_b}{c_t}$ is required to operate the device with a fixed filling factor $\nu_{EW}$ in the fractional quantum Hall regime. The thickness of the hBN dielectric spacers used to fabricate the device are not precisely equal - rather than calculating the ratio $\alpha$ from distances measured in AFM topography, from which the top and bottom hBN thickness are known to be $x$ and $y$ nm, it is desirable to measure the ratio in transport to ensure that all stray capacitive effects are accounted for, to avoid deviations from the expected capacitance to each gate layer causing any uncertainty about the bulk filling factor in any region of the device.

In the quantum Hall regime a number of conductance features depend only on the electron density of the 2DEG, so tracking the shift of these features as voltages on gates in the top and bottom layer are changed allows us to determine $\alpha$. For example, we can measure the shift in top gate voltage $\Delta V_t$ of the center of a known feature in the conductance when the bottom gate voltage is changed by $\Delta V_b$. Since we assume the conductance peak appears at a fixed density n,

$$
n = c_{t}V_{ti} + c_{b}V_{bi} = c_{t}V_{tf} + c_{b}V_{bf}
$$

we can use the measured value of this shift to calculate $\alpha$:

$$
\alpha = \frac{c_b}{c_t} = - \frac{\Delta V_t}{\Delta V_b}
$$

Here we make use of the fact that the longitudinal conductance $\sigma_{xx}$ of a FQH fluid vanishes when the Hall conductance $\sigma_{xy}$ is on a plateau and exhibits a peak between Hall plateaus to determine $\alpha$. In particular, we measure the Hall and longitudinal conductance as a function of $V_W$ on the W side of the device at B=10T for two back-gate voltages, $V_{bg}$ = $1.0V$ and $2.0V$, and fit the shift $\Delta V_W$ between the two traces. This value is extracted from the features around the $\nu = -2/3$ plateau in particular since it shows the least qualitative change between each trace among the various discernible features in $\sigma_{xx}$ (due to the sensitive dependence of the edge state equilibration on the doping of the single-gated region of each contact, which varies as the W gate value is changed, there are noticeable qualitative distortions of the conductance comparing the two traces).

The shift $\Delta V_W$ is numerically determined by minimizing the value of $X$, the norm-squared difference between the two $\sigma_{xx}$ traces with respect to $\Delta V_W$:

$$
X = \int \Big|\sigma_{xx}(V_{bg}=1,V_W=V)-\sigma_{xx}(V_{bg}=2,V_W=V+\Delta V_W)\Big|^2 dV
$$

For $\Delta V_{bg} = 1.0V$, we calculate a shift $\Delta V_W=-1.773V$, so $\alpha = 1.773$.

\section{Statistics of Fractional Quantization Plateaus in Bulk Integer Filling}

\begin{figure}
    \centering
    \includegraphics[width = 90mm]{supp_fractional_reconstruction_histogram.pdf}
    \caption{Histogram of diagonal conductance values occurring in Figure 3c of the main text}
    \label{fig:supp_histogram}
\end{figure}

The diagonal conductance map of the $\nu_{EW} = -1$ state at B=13T presented in the main  text (Figure \ref{fig:fractional_qpc}c) shows a number of remarkable features between $G_D = 1$ and $G_D = 0$. The value of each plateau is not entirely clear from the full 2-dimensional plot since the quantization is not precise and varies slightly with the exact electrostatic profile near the saddle point - however, a histogram of the statistical frequency of conductance values across the entire plot, shown in Figure \ref{fig:supp_histogram}, reveals a number of sharp peaks corresponding to conductance plateaus between 0 and 1. The most prominent occurs at $G_D = 2/3$. A broader peak is observed near 1/3, though the quantization is less exact, and several further sharp peaks are seen centered near 0.2, 0.5, and 0.8. The origin of these peaks is unknown, and may be due to a more complicated reconstructed edge structure than is accounted for by any model discussed in this work.

\section{Filling Factor Maps at 6T}
In order to appropriately interpret the integer QPC operation described in the main text, it is imperative to know the exact mapping between applied gate voltages and the bulk filling factor in the east and west regions. Figure-\ref{fig:filling_factor_maps}a shows a measurement of $1/R_{xy}$ on the west side of the device, which shows quantized plateaus at integer filling factors up to $\nu = 6$.  $1/R_{xy}$ here is measured as in figure-1f of the main text.   The inset in figure-\ref{fig:filling_factor_maps}a additionally shows the full width of the correlated insulating gap at $\nu = 0$ directly by observing the measured current through the west region drop to zero.  Given that we are able to measure the capacitance ratio precisely the filling factor map can be extrapolated to any back gate voltage.  Figure-\ref{fig:filling_factor_maps}b shows the same $1/R_{xy}$ dataset but taken at $V_{bg} = -0.5V$.  Focusing on the regions between $\nu = 0$ and $\nu = 2$, clear pleateus can be observed at 1/3, 2/3, 5/3, 3/5, and 8/5.  This is an additional demonstration that the edge mode equilibration is sensitive to the contact doping.  Adjusting the bottom gate requires a concomitant change in the top gate voltage which also dopes part of the contacts, leading to a clearer emergence of FQH plateaus at 6T.

\begin{figure}[h!]
    \centering
    \includegraphics[width = 90mm]{sup_fig_fillingfactor_maps_6T.pdf}
    \caption{\textbf{Filling factor maps at B = 6T} (a) $1/R_{xy}$ measured on the west side of the device versus $V_{W}$ while $\nu_{N} = \nu_{E} = \nu_{E} = 0$ is kept fixed, and $V_{bg} = 0.5V$.  Inset: current measured during $1/R_{xy}$ sweep showing $\nu = 0$ gap.  (b) $1/R_{xy}$ measured as in (a) but with $V_{bg} = -0.5V$, here the edge state equilibration is improved and clear fractional plateaus at various values of $n/3$ and $n/5$ can be observed. }
    \label{fig:filling_factor_maps}
\end{figure}

\section{Extended Integer QPC Operation}
\begin{figure*}
    \centering
    \includegraphics[width = 180mm]{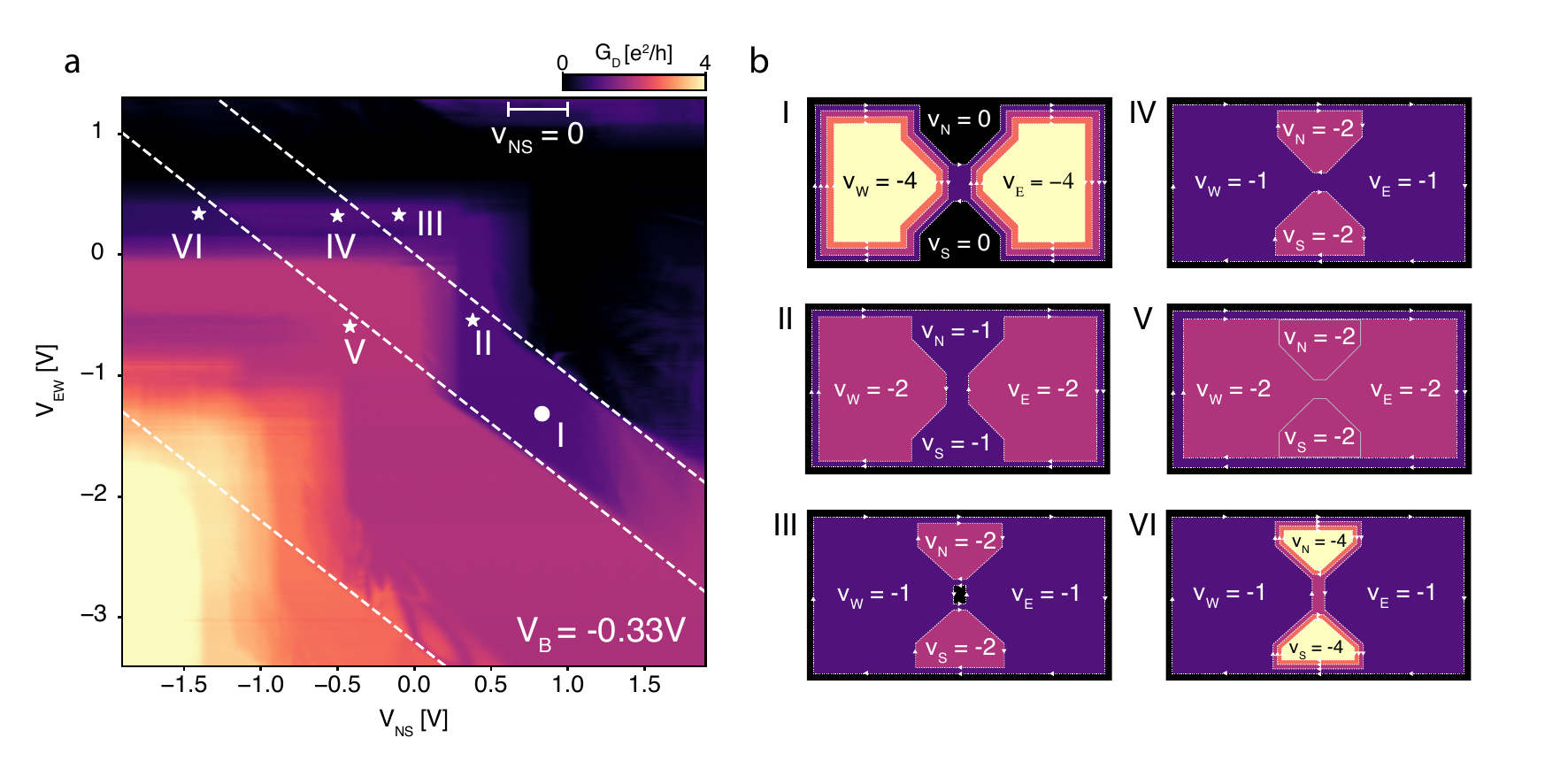}
    \caption{\textbf{Extended Integer QPC Operation} (a) $G_D$ plotted versus $V_{NS}$ and $V_{EW}$ with $V_{bg} = -0.33V$.  Lines parallel to $V_{NS} - V_{EW} = 0$ demarcate transitions between integer $G_D$ values indicating a rapid change in the density at the QPC.  Roman numerals I-VI correspond to various unique combinations of $\nu_{NS}$, $\nu_{EW}$, and $\nu_{qpc}$, where $\nu_{qpc}$ is the filling factor at center of the QPC.  (b) Illustrations of the filing factor in each region of the device corresponding to the operating points in (a).}
    \label{fig:ext_integer_qpc}
\end{figure*}

The full parameter space that determines the value of $G_D$ when tuned with $V_{NS}$ and $V_{EW}$ is complex and deserves an exhaustive treatment.  To get oriented, point I in figure-\ref{fig:ext_integer_qpc}a corresponds to point II in figure-2b of the main text.  Here $\nu_{NS} = 0$ forming a narrow constriction in a bulk $\nu_{EW} = -4$, only allowing 1 edge mode to traverse the QPC giving $G_D = 1$.  The lines that are perpendicular to $V_{NS} + V_{EW} = 0$ (indicated as dashed white lines in figure-\ref{fig:ext_integer_qpc}a) are interpreted as boundaries between operating points that have differing values of $\nu_{qpc}$, the filling factor in the center of the qpc.  Consider the situation where point I on figure-\ref{fig:integer_qpc}b were adjusted upward (keeping $\nu_{NS} = 0$) such that it crosses the upper dashed line so $G_D = 1 \rightarrow 0$.  Given that the edge modes only traverse the device via the QPC in this regime, the sudden change in $G_D$ can only be explained by a corresponding change in filling factor in the same location.  A similar argument can be run by adjusting point I downward such that $G_D = 1 \rightarrow 2$.  The filling factor in the center of the QPC is determined via fringe fields from the four quadrant gates when $V_{bg}$ is fixed, and due to the high degree of symmetry of the AFMLAO cut gate structure the filling factor in the QPC center can be kept constant by adjusting $V_{EW}$ in tandem with an equal but opposite change in $V_{NS}$.  It is clear that there are distinct diagonal lines in figure-\ref{fig:ext_integer_qpc}a when   It is expected that these electrostatics remain even when $G_D$ is determined primarily by effects outside of the QPC.  Consequently, we expect the white dashed lines which separate regions of differing $\nu_{qpc}$ to extend throughout the dataset.

With this in hand, it is possible to determine $\nu_{EW}$, $\nu_{NS}$, and $\nu_{qpc}$ for each point marked in figure-\ref{fig:ext_integer_qpc}a.  All of the filing factors for each point in figure-\ref{fig:ext_integer_qpc}a will be listed in a table at the end of this section.  However point VI is of particular note since it falls to the left of a boundary in $G_D$, where $G_D$ falls to some fractional value below 1, that is intersected by the a dashed line making the transition indicative of qpc physics.  The corresponding filling factors there are $\nu_{NS} = -4$, $\nu_{EW} = -1$, and $\nu_{qpc} = -2$ since IV falls between the two lowest dashed lines.  Given that $\nu_{EW} = -1$, $\nu_{qpc} = -2$ is the first filled Landau level which introduces and edge state that directly shorts the device from the north to south regions via the QPC.  The formation of such a $pp'p$ junction has been extensively studied, and has been shown to lead to a reduction in $G_D$ \cite{zimmermann_tunable_2017, deprez_tunable_2021}. 

\begin{center}
\begin{tabular}{||c c c c c||}
 \hline
 \textbf{Index} & $G_D [e^2/h]$ & \textbf{$\nu_{EW}$} & \textbf{$\nu_{NS}$} & \textbf{$\nu_{qpc}$} \ \\ [0.8ex] 
 \hline\hline
 I & 1 & -4 & 0 & -1 \\ 
 \hline
 II & 1 & -2 & -1 & -1 \\
 \hline
 III & 1 & -1 & -2 & 0 \\
 \hline
 IV & 1 & -1 & -2 & -1 \\
 \hline
 V & 2 & -2 & -2 & -2 \\ 
 \hline
 VI & 0.7 & -1 & -4 & -2 \\ [1ex] 
 \hline
\end{tabular}
\end{center}

\section{A Comparison of Reconstructed Edge Models}

\begin{figure*}[ht]
    \centering
    \includegraphics[width = 180mm]{supp_mode_structure.pdf}
    \caption{\textbf{Reconstructed Models of the Edge at $\nu=-1$ and $\nu = -5/3$}}
    \label{fig:supp_mode_structure}
\end{figure*}

\section{References}
\bibliographystyle{unsrt}
\bibliography{references}